\documentclass{imamat}

\jno{drnxxx}
\usepackage{amsmath} 
\usepackage{amsthm} 
\usepackage{graphicx} 
\usepackage{subfig,color} 

\usepackage{booktabs}
\usepackage{longtable}

\newcommand{\kcpf}{k_{\text{\scriptsize cPF}}}
\newcommand{\kcHopf}{k_{\text{\scriptsize cHopf}}}
\DeclareMathOperator{\sech}{sech}
    
     \newcommand{\RN}[1]{%
\textup{\uppercase\expandafter{\romannumeral#1}}%
}
    
\begin{document}

\title{Spatial localisation beyond steady states in the neighbourhood of the Takens--Bogdanov bifurcation}

\shorttitle{Spatial localisation beyond steady states} 
\shortauthorlist{Alrihieli, Rucklidge and Subramanian} 

\author{
\name{Haifaa Alrihieli}
\address{School of Mathematics, University of Leeds, Leeds LS2 9JT, United Kingdom}\\
Faculty of Science, University of Tabuk, Tabuk 47512, Saudi Arabia
\name{Alastair Rucklidge}
\address{School of Mathematics, University of Leeds, Leeds LS2 9JT, United Kingdom}
\and
\name{Priya Subramanian$^*$}
\address{Mathematical Institute, University of Oxford, Oxford OX2 6GG, United Kingdom}
\email{$^*$Corresponding author: priya.subramanian@maths.ox.ac.uk}}

\maketitle

\begin{abstract}
{The coincidence of a pitchfork and Hopf bifurcation at a Takens--Bogdanov bifurcation occurs in many physical systems such as double-diffusive convection, binary convection and magnetoconvection. Analysis of the associated normal form, in one dimension with periodic boundary condition, shows the existence of steady patterns, standing waves, modulated waves and travelling waves, and describes the transitions and bifurcations between these states. Values of coefficients of the terms in the normal form classify all possible different bifurcation scenarios in the neighbourhood of the Takens--Bogdanov bifurcation \citep{dangelmayr1987takens}. In this work we develop a new and simple pattern-forming PDE model, based on the Swift--Hohenberg equation, adapted to have the Takens--Bogdanov normal form at onset. This model allows us to explore the dynamics in a wide range of bifurcation scenarios, including in domains much wider than the lengthscale of the pattern. We identify two bifurcation scenarios in which coexistence between different types of solutions is indicated from the analysis of the normal form equation. In these scenarios, we look for spatially localised solutions by examining pattern formation in wide domains. We are able to recover two types of localised states, that of a localised steady state in the background of the trivial state and that of a spatially localised travelling wave in the background of the trivial state which have previously been observed in other systems. Additionally, we identify two new types of spatially localised states: that of a localised steady state in a modulated wave background and that of a localised travelling wave in a steady state background. The PDE model is easy to solve numerically in large domains and so will allow further investigation of pattern formation with a Takens--Bogdanov bifurcation in one or more dimensions and the exploration of a range of background and foreground pattern combinations beyond steady states.
}
{Pattern formation, Takens--Bogdanov bifurcation, Spatial localisation}\\
2000 Math Subject Classification: 35B22, 35B36, 37G05, 37N10, 37Mxx
\end{abstract}

\section{Introduction} \label{sec:Introduction}

The Takens--Bogdanov bifurcation exhibits a variety of dynamical behaviours. The bifurcation occurs when a Hopf and pitchfork bifurcation coincide, and arises in physical problems such as double-diffusive convection in a horizontal layer of fluid heated from below \citep{rucklidge1992chaos}. This system has two competing gradients that drive motion in the fluid:  the temperature gradient and the solute gradient. With low solute gradient, the first bifurcation from the resting (trivial) state as the temperature gradient is increased is a pitchfork bifurcation leading to steady convection. With larger solute gradient, the bifurcation changes to a Hopf bifurcation leading to oscillatory convection. In double-diffusive convection with idealised boundary conditions, these two forms of convection set in with the same horizontal wavelength \citep{veronis1965finite}. In two dimensions with periodic boundary conditions the point where the pitchfork and Hopf bifurcation coincide is called the Takens--Bogdanov point. The normal form that describes the amplitude close to the Takens--Bogdanov (TB) point for a system with O(2) symmetry is \citep{dangelmayr1987takens}
 
\begin{equation}\label{eq:1.1}
\ddot{z}= \mu z +  A |z|^2 z +\epsilon \left( \nu \dot{z}   +C \left(\dot{z} \bar{z}+ z \dot{\bar{z}}\right) z +   D |z|^2 \dot{z} \right) + O(\epsilon^2),  \qquad \epsilon \ll 1
\end{equation}
where $z$ is the complex amplitude of the pattern, $\mu$ and $\nu$ are the unfolding parameters,  $A,C$ and $D$ are constants, the dot denotes  differentiation with respect to time, and $\epsilon$ controls how close the system is to the TB point.

Different bifurcation scenarios obtained by the analysis of the amplitude equation are found close to onset \citep{dangelmayr1987takens,knobloch1986oscillatory}, with several different types of patterns: steady states (SS), travelling waves (TW), standing waves (SW) and modulated waves (MW). In domains that are many times wider than the preferred wavelength, extended TW, SW, and MW solutions have been found in numerical investigations of the partial differential equations (PDEs) for thermosolutal convection \citep{deane1988traveling,spina1998confined,turton2015prediction}. In a similar scenario, in binary convection, the application of a thermal gradient to a mixture sets up a competing concentration gradient due to the Soret effect. In this system, a transition from SS to TW has been observed in numerical simulations \citep{zhao2015numerical,barten1995convectionE}, while a nonlinear SW solution has been numerically obtained by \citet{matura2004standing} and \citet{jung2004oscillatory}.

In addition to patterned states that are spatially extended, i.e., span the entire domain, parameter values where both the trivial state and a periodic SS state are both dynamically stable allow for the existence of spatially localised states. In the subcritical regime with coexistence between the trivial state and the periodic SS branches, spatially localised steady states (LSS) in a background of the trivial state undergoing homoclinic snaking have been obtained in numerical investigations of  thermosolutal convection \citep{beaume2011homoclinic} and binary convection  \citep{batiste2006spatially,mercader2009localized}. The snaking branches behave like those familiar from the Swift--Hohenberg equation \citep{burke2007snakes}. At a given  Rayleigh number, odd and even branch solutions with different number of rolls can be found. 

For binary convection, the system undergoes a subcritical Hopf bifurcation to  TW for negative separation ratio \citep{zhao2015numerical}. 
In the parameter regime where the TW  bifurcate subcritically  from the conduction state, localized travelling waves (LTW)  have also been  obtained. The LTW solution refers to the spatially localized cells whose envelope moves with a characteristic speed in a background of the trivial state.  In contrast to LSS, the LTW have  fixed and uniquely selected width, which was discovered in experimental  \citep{kolodner1994stable,kolodner1991stable,niemela1990localized,kolodner1991collisions} and numerical \citep{barten1991localized,taraut2012collisions,barten1995convection} studies of binary convection, with a negative separation ratio $=-0.08$. This was also observed later in numerical simulations of the full system of binary convection with different but still small negative separation ratios of $-0.123$ \citep{watanabe2012spontaneous} and $-0.1$ \citep{zhao2015numerical}.

In this paper, we develop a new and simple model as a useful description of the qualitative features of double-diffusive convection. Our model is a PDE based on the Swift--Hohenberg equation but adapted to have the TB normal form at onset. This allows for an exploration of the dynamics of localised steady and time dependent patterns in very wide domains. Our model can access most of the bifurcation scenarios that occur in the TB normal form and so it is relevant to other pattern-forming systems whose dynamics can be reduced to a TB normal form.The model recovers LSS and LTW as documented above, as well as two new localised patterns: LSS in an oscillating background and LTW moving through a background of SS. 

 In Section \ref{sec:Derivation of the model}, we develop the linear part of the model  by reproducing the dynamics of double-diffusive convection.  In Section \ref{sec:Nonlinear part}, we discuss the nonlinearities which we can add to the model, taking into account Lyapunov stability. In Section \ref{sec:TB},  the model is reduced to the Taken--Bogdanov normal form by applying a weakly nonlinear analysis. In Section \ref{sec:Relate the model to  DK cases}, we identify parameter combinations in the model at which we can observe different dynamical behaviours close to the TB bifurcation. In Section~\hbox{\ref{sec:lss}}, we obtain localized SS with trivial state background and localized SS with SW background using time stepping and observe snaking in the branch of localized SS with trivial state background using continuation. Localized TW with trivial state and SS background are discussed in Section \ref{sec:ltw} using time stepping. We conclude in Section \ref{sec:conclusions}.


\section{Designing the linear dynamics near the TB bifurcation} \label{sec:Derivation of the model}

\begin{figure}
\centering{\includegraphics[width=0.9\linewidth]{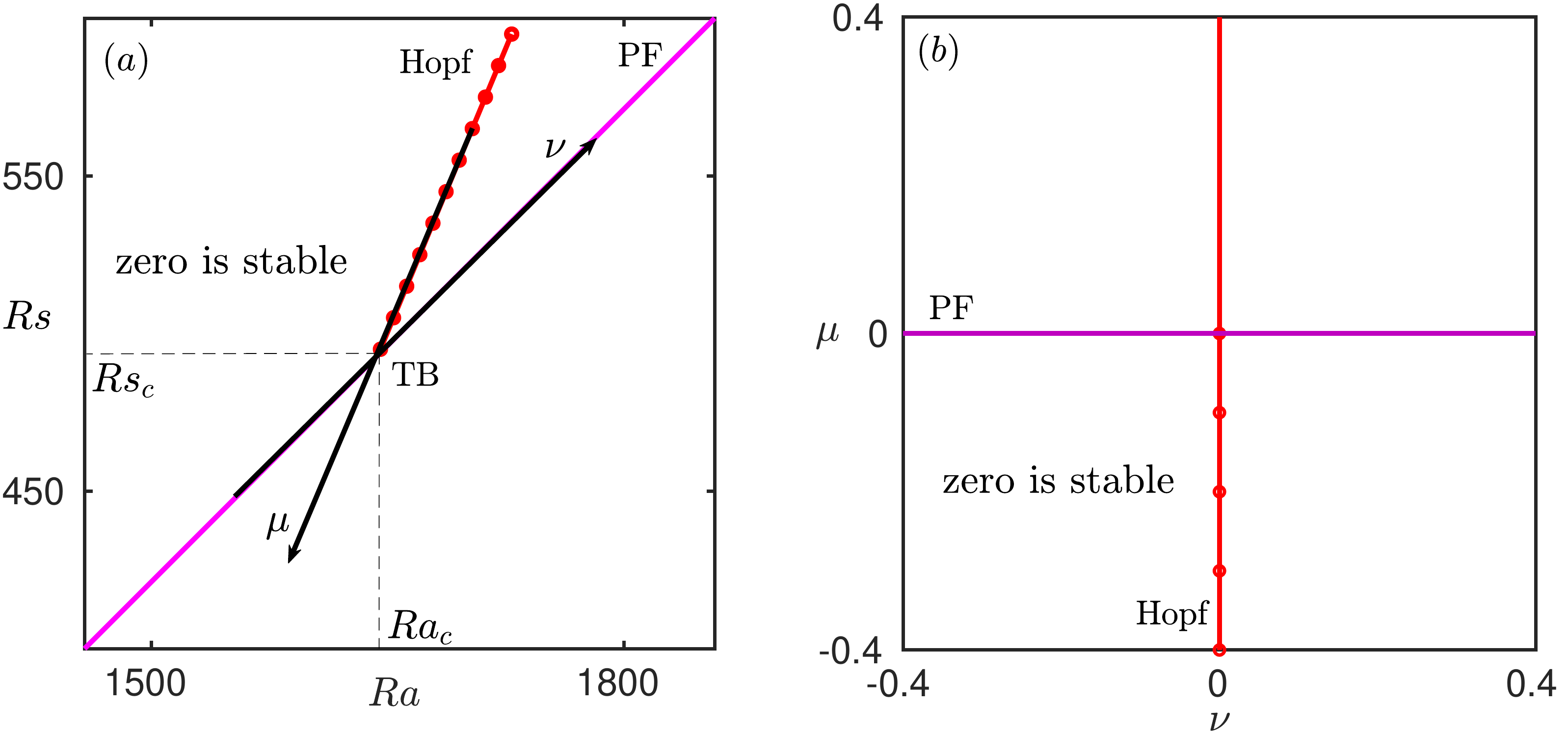} }
     \caption{(a) Unfolding diagram for the pitchfork (pink solid line) and Hopf (red line with circle markers) in the $(Ra, Rs)$-plane of double-diffusive convection.  At the co-dimension two Takens--Bogdanov point (TB) with  $(Ra,Rs) =(Ra_c, Rs_c)=(1647, 495)$, the Hopf and pitchfork bifurcations are coincident. For each $Rs$, the quiescent/zero state is stable at small values of $Ra$ till we cross either a pitchfork/Hopf transistion as we increase $Ra$. (b) The $(\nu,\mu)$-plane mapped from (a). }
    \label {fig:1} 
    \end{figure}

The first part of designing a model system that has a Takens--Bogdanov (TB) bifurcation requires the possibility for both a pitchfork and a Hopf bifurcation. We build a minimal model for the TB bifurcation by reproducing the dynamics of double-diffusive convection, starting with the design of the linear dynamics. Two different density gradients drive the dynamics in double-diffusive convection: thermal gradients quantified with a thermal Rayleigh number $Ra$ and solutal gradients quantified with a solutal Rayleigh number $Rs$. The stable quiescent state in the system becomes unstable with increasing thermal gradients and starts to convect. When $Rs$ is less than a critical value $Rs_c,$  the quiescent state undergoes a pitchfork bifurcation leading to steady convection as the temperature gradient $Ra$ increases. At large solutal gradients with $Rs>Rs_c$, this behaviour changes and the quiescent state loses stability via a Hopf bifurcation leading to oscillatory convection. The point where the primary bifurcation changes from pitchfork to Hopf bifurcation with $(Ra, Rs)=(Ra_c, Rs_c)$ is called the Takens--Bogdanov point, as shown in Fig.\,\ref{fig:1}(a). 

In order to replicate this behaviour, we use two control parameters $\nu$ and $\mu$ in the new model, where the change of sign in $\nu$ corresponds to the loci of pitchfork bifurcations and the change of sign in $\mu$ (with $\nu<0$) corresponds to the loci of Hopf bifurcations, as shown in Fig.\,\ref{fig:1}. Such an identification allows us to decompose the behaviour close to the TB point as follows. Starting from large negative values, variations of parameters above the diagonal $\mu=\nu$, as shown in Fig.\,\ref{fig:1}($b$), allow for the occurrence of pitchfork bifurcation, while parameter variations below this diagonal causes a Hopf bifurcation to occur, followed by a pitchfork bifurcation. In this way we are able to replicate the different bifurcation scenarios from double-diffusive convection. 

\begin{figure}  
\centering{\includegraphics[width=0.9\linewidth]{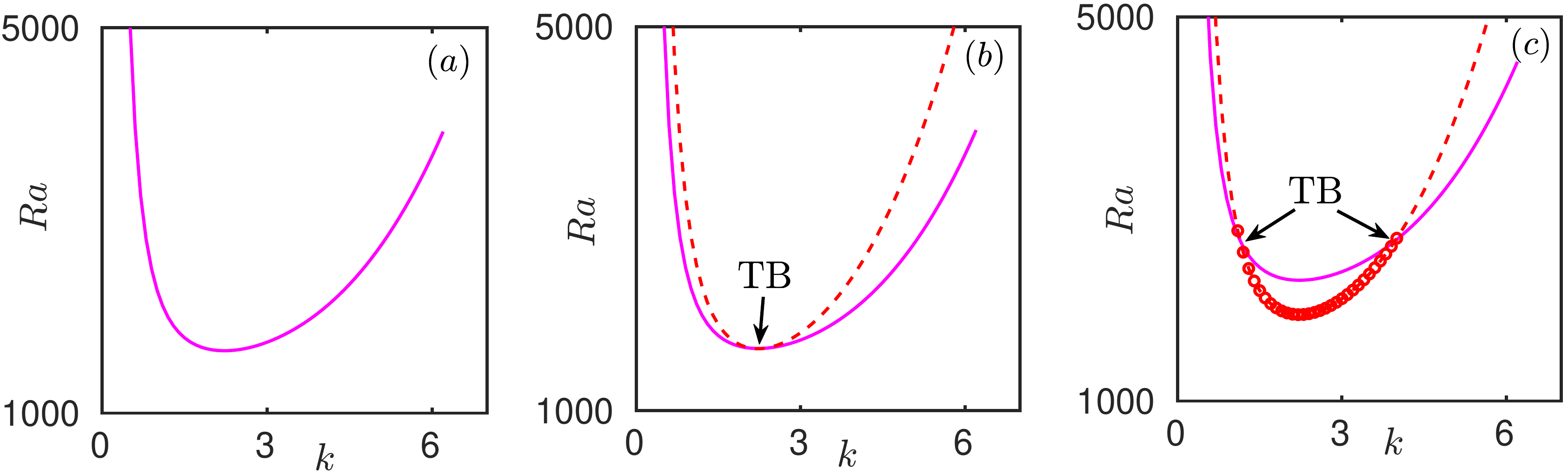} }
\caption{Plot of the neutral stability curves for pitchfork and Hopf bifurcation for double-diffusive convection. The pink solid line refers to the loci of pitchfork bifurcation and the red dashed line refers to the loci of Hopf bifurcation. The red circle markers identify locations where the Hopf bifurcation is the primary bifurcation. ($a$)  $Rs< Rs_c$ where pitchfork bifurcations are the primary bifurcation, ($b$) $Rs= Rs_c=495$ where the pitchfork and Hopf bifurcation are coincident at a TB point, and ($c$) $Rs> Rs_c$ where the pitchfork and Hopf bifurcation can be primary bifurcations. In this case two TB points can be identified.}
\label {fig:2}
\end{figure}

The second factor to include is the variation of the linear stability threshold with wavenumber $k$. In the case of double-diffusive convection, the linear stability thresholds for pitchfork and Hopf bifurcations are as shown in Fig.\,\ref{fig:2}. In all three cases, we observe that both the pitchfork and Hopf thresholds vary with the wavenumber $k$. For example, the threshold for pitchfork bifurcations (shown in pink) has a minimum $Ra = Ra_0$ at $k=k_c= \frac{\pi}{\sqrt{2}}$. Near this critical wavenumber $k_c$, the critical Rayleigh number varies with the square of the distance of the current wavenumber $k$ from $k_c$. This means that close to $k=k_c$, $Ra_0$ can be expanded as
\begin{equation} 
Ra_0(k^2) \approx Ra_0 +  \frac{Ra_0''}{8 k_c^2} (k_c^2- k^2)^2, 
\end{equation} 
where $Ra_0''= {d^2 Ra_0}/{d k^2}$. In order to reproduce this behaviour, we incorporate such a variation into the two linear stability thresholds by writing $\mu$ and $\nu$  in terms of $(k_c^2-k^2)^2$ as 
\begin{equation} 
\begin{aligned} 
\mu &=  (\kcpf^2 - k^2)^2,\\
\nu &=  b(\kcHopf^2 - k^2)^2\,.  
 \label{eq:2.10}
\end{aligned}
\end{equation}
Here the constant $b$ is used to change the rate at which Hopf bifurcation varies with $(k_c^2-k^2)^2$. From the case of double diffusion shown in Fig.\,\ref{fig:2}, we see that we require $b>1$. Other choices can be made depending on the specific system of interest. 

Close to the TB point, the PDEs that govern small amplitude thermosolutal convection can be reduced to the linear second order Van der Pol--Duffing equation \citep{veronis1965finite}
  \begin{equation} 
  \ddot{u}- \kappa \dot{u} - \lambda u = 0, 
    \label{eq:2.12}
  \end{equation}
where $u$ is the amplitude of the lowest-order mode of the stream function, $\kappa, \lambda $ are unfolding parameters and the dot indicates the derivative with respect to time. We start with this dynamical equation and look to fit in our two parameters in terms of the unfolding parameters in this model below. 
  
For the linear equation in (\ref{eq:2.12}), we can assume a solution of the form $ u= e^{\sigma t}, $ where $\sigma$ is the growth rate. The characteristic equation then takes the following form 
  \begin{equation} 
  \sigma^2 - \kappa \sigma -\lambda = 0.
  \end{equation} 
  This is equivalent to an eigenvalue problem for $\sigma$ of the form 
  	\begin{equation}\label{eq:2.13}
	\sigma \left[ {\begin{array}{c}
		1	\\
			\sigma\\ 
		\end{array} } \right]=
	\	\left[ {\begin{array}{cc}
0& 1\\
		\lambda& \kappa \\
		\end{array} } \right]
	\left[ {\begin{array}{c}
	 1	\\
	 \sigma\\ 
		\end{array} } \right]
\equiv	\bf{L} 	\left[ {\begin{array}{c}
	 1	\\
	 \sigma\\ 
		\end{array} } \right].
\end{equation}
 A pitchfork bifurcation (with one zero eigenvalue) occurs when the determinant is zero, i.e., $\lambda =0,$ and a Hopf bifurcation (with purely  imaginary eigenvalues) occurs when the trace is zero and the determinant is positive, i.e., $ \kappa=0 $ and $\lambda <0$.  At the TB point, the system has two zero eigenvalues, i.e. $\kappa = \lambda =0$.
We can relate $\mu$ and $\nu$ from Eqn. (\ref{eq:2.10}) to this two dimensional dynamical system by setting
\begin{equation} 
\begin{aligned} 
\lambda &= \mu-  (\kcpf^2 - k^2)^2,\\
\kappa& = \nu - b (\kcHopf^2 - k^2)^2\,.
   \label{eq:2.14}
\end{aligned} 
\end{equation}
This gives the relations that $\lambda=0$ when $\mu= (\kcpf^2 - k^2)^2$, and $\kappa =0$ when $\nu= b (\kcHopf^2 - k^2)^2$. This implies that the linear operator in terms of Eqn. (\ref{eq:2.14}) is then given as
\begin{equation} 
\bf{L}=	\	\left[ {\begin{array}{cc}
0&1 \\
		\mu-  (\kcpf^2 - k^2)^2& \nu - b (\kcHopf^2 - k^2)^2 \\
		\end{array} } \right].
		 \label{eq:2.15}
\end{equation}
Consequently, we can write  the linear equation (\ref{eq:2.12}) as 
  \begin{equation} 
  \ddot{u}- \left(\nu - b (\kcHopf^2 - k^2)^2\right) \dot{u} -\left(\mu-  (\kcpf^2 - k^2)^2\right) u = 0, 
    \label{eq:2.16}
  \end{equation}
  where $u(t)$ is now the amplitude at lowest order of the mode $e^{ikx}.$ This equation in Fourier space can be converted into a PDE by replacing $k^2$ with $- {\partial^2}/{\partial x^2} $ and considering $u$ to be a function of $x$ and $t$. 
The ODE (\ref{eq:2.16}) converted to a PDE is
\begin{equation} \label{eq:2.17}
u_{tt} -\left(\nu-b\left(\kcHopf^2+\frac{\partial^2}{\partial x^2} \right)^2\right) u_t - \left(\mu-\left(\kcpf^2+\frac{\partial^2}{\partial x^2}\right)^2\right) u=0,
\end{equation}
 where $u_{tt} = {\partial^2 u}/{\partial t^2}$ and $u_{t} = {\partial u}/{\partial t}$ are the  derivatives with respect to time, and the parameters $\kcpf$ and $\kcHopf$  are the wavenumbers for the pitchfork and Hopf bifurcations.
 
The general choice of $\kcpf \neq \kcHopf$, is relevant to problems where the pitchfork and Hopf bifurcation have different critical wavenumbers, e.g., magnetoconvection \citep{chandrasekhar2013hydrodynamic,arter1983nonlinear, weiss1981convection,proctor1982magnetoconvection} and rotating convection \citep{clune1993pattern,veronis1966motions, veronis1968large, zimmermann1988effect}. Since we are interested in thermosolutal convection, where the pitchfork and Hopf bifurcations have the same critical wavenumbers, for this paper we let $\kcpf=\kcHopf=1$.  Then the linear second order partial differential equation that models the dynamics at small amplitudes near a TB point takes the following from:  
\begin{equation} \label{eq:2.18}
u_{tt}  = \left(\nu-b\left(1+\frac{\partial^2}{\partial x^2} \right)^2\right) u_t + \left(\mu-\left(1+\frac{\partial^2}{\partial x^2}\right)^2\right) u \equiv \mathcal{M}_1 u +\mathcal{M}_2 u_t=0.
\end{equation}
The dispersion relation can be determined by studying the eigenvalues of the model given by 
\begin{equation} 
\sigma(k) = \frac{\left(\nu-b\left(1-k^2\right)^2\right) \pm \sqrt{\left(\nu-b\left(1-k^2\right)^2\right)^2 + 4\left(\mu-\left(1-k^2\right)^2\right)}}{2}. 
\label{eq:3-3-4}
\end{equation}
The growth rate $\sigma$ is a function of the wavenumber $k$, so some modes could decay and while others can grow. If the eigenvalues for all $k$ have negative real parts the evolution decays and the zero solution is linearly stable. If any eigenvalue for any $k$ has positive real part the evolution grows and the zero solution is linearly unstable.  


\section{Selection of the nonlinear terms} \label{sec:Nonlinear part} 

The model given in Eqn.\,(\ref{eq:2.18}) is a second order linear partial differential equation in time which has been designed to reproduce the linear stability results of double-diffusive convection. This section deals with determining the choice of nonlinear terms in the model, with two aims: first, to have a globally stable nonlinear dynamical system and second, to be able to access the variety of dynamical behaviours that can occur in the neighbourhood of the TB bifurcation. 

We consider only those nonlinearities which are invariant with respect to reflection about the $x$-axis (to replicate this symmetry in double-diffusive convection) and include nonlinearities both in the field $u$ and its time derivative $u_t$. We can additionally classify the nonlinear terms in the order at which they contribute to the dynamics. For example, quadratic nonlinear terms can include 
 \begin{equation}
 u^2, u u_t, u_t^2, u u_{xx},  (u_x)^2, u u_{txx}, u_t u_{xx}, u_x u_{tx}, 
 \label{eq:3.1}
 \end{equation}
 while cubic nonlinearities can include terms such as 
 \begin{equation} 
  u^3,u u_t^2, u^2 u_t, u_t^3, (u u_{xx}) u_t,   (u_x)^2 u_t,  u^2 u_{txx}, u u_t  u_{xx}, u u_x u_{tx}\,.
   \label{eq:3.2}
  \end{equation}
Similar terms (without the $u_t$ contributions) have been incorporated into various generalisation of the Swift--Hohenberg equation \citep{burke2012localized, kozyreff2007nonvariational, crawford1999oscillon}. 
Including all the nonlinearities mentioned would make the model very complicated. We consider the two criteria identified at the beginning of this section and proceed to create a candidate nonlinear model of the form, 
 \begin{equation} 
\frac{\partial^2 u}{\partial t^2}  = \mathcal{M}_1 u + \mathcal{M}_2 u_t +Q_1u^2 +C_1u^3 +C_2 u^2 u_t +C_3 u_t^3\,.
\label{eq:3.3}
\end{equation}
In this model we have chosen simple polynomial nonlinearities in the form of $u^2$, $u^3$ and $u_t^3$ along with the additional mixed nonlinear term $u^2 u_t$, each with a constant coefficient. 

The first step is to ensure global stability during evolution. In order to test this, we look to determine a Lyapunov function for the dynamics in the spirit of the Swift--Hohenberg equation. With this in mind, we consider the Lyapunov functional
\begin{equation} \label{eq:3.7}
\mathcal{F}(t)=\int_{0}^{L}  \left[ \frac{1}{2} u_t^2 - \frac{1}{2} \mu u^2 + \frac{1}{2}\left[\left(1+\frac{\partial^2}{\partial x^2}\right)u\right]^2 - \frac{1}{3}Q_{1}u^3 - \frac{1}{4} C_{1} u^4 \right] dx.
\end{equation}	
where $L$ is the length of the periodic domain. Our aim is to show that this function is bounded below and decreases with time for large $u$ and $u_t$. Having $C_1<0$ is sufficient for $\mathcal{F}(t)$ to be bounded from below. Differentiating $\mathcal{F}(t)$ with respect to time we have 
\begin{align}  \label{eq:3.12}
\frac{ d\mathcal{F}}{ dt}&=  \int_{0}^{L}  \left[ u_{tt} - \mu u  + \left(1+\frac{\partial^2}{\partial x^2}\right)^2 u   - Q_{1}u^2-C_{1}  u^3 \right] u_t\, dx\\
 &= \int_{0}^{L}  \left[ \nu u_t -b \left(1+\frac{\partial^2}{\partial x^2}\right)^2 u_t+ C_{2} u^2 u_t + C_3 u_t^3\right] u_t\, dx. 
\end{align} 
If $u_t=0$ for all $x$, then we see that the above relation vanishes with ${ d\mathcal{F}}/{ dt}=0$ and we have an equilibrium. 
Now, we consider a non-equilibrium point in the dynamics with large non-zero values of $u$ and $u_t$ and want show that the rate of change of $\mathcal{F}$ is still negative. In order to do this, we can simplify the expression above a bit further. At large $u$ and $u_t$, the last two quartic terms dominate, so we write them as follows:
\begin{equation}  \label{eq:3.14}
T = \int_{0}^{L} \left[ C_{2} u^2 u_t^2 + C_3 u_t^4\right] dx. 
\end{equation}
We can re-cast the states into an amplitude-phase form with
\begin{equation}  \label{eq:3.15}
u(x,t) = R \cos \phi, \qquad \textrm{and} \qquad   u_t(x,t) = R \sin \phi, 
\end{equation}
 where $R(x,t)$ is a large radius and $\phi(x,t)$ is an angle. Substituting (\ref{eq:3.15}) into (\ref{eq:3.14}) and simplifying, we get  
\begin{equation}
T  = \int_{0}^{L}  \left[R^4 \left(\frac{1}{2}\left(C_{2}  + C_{3} \right)+\frac{1}{2} \left(C_{2}  - C_{3} \right) \cos 2\phi 
\right) \sin^2\phi \right] dx.
\label{eq:3.16}
\end{equation}

If trajectories are to remain bounded for any choice of $\nu$, ${d\mathcal{F}}/{dt}$ must be negative for any $(u,u_t)$ large enough as long as $u_t$ is not zero for all $x$. This is guaranteed if $T<0$ for any $\phi$  as long as $\sin \phi$ is not zero for all $x$, which requires 
\begin{equation}
C_{2}  + C_{3}<0 \qquad \textrm{and} \qquad 0< \sqrt{\left(\frac{1}{2} \left(C_{2}  - C_{3}\right)\right)^2 } <-\frac{1}{2}\left(C_{2}  + C_{3} \right) \implies C_2 C_3 >0\,.
\end{equation}
Therefore the model second order partial differential equation takes the following form 
 \begin{equation} 
\frac{\partial^2 u}{\partial t^2}  = \left(\mu-\left(1+\frac{\partial^2}{\partial x^2}\right)^2\right) u+\left(\nu-b\left(1+\frac{\partial^2}{\partial x^2}\right)^2\right) \frac{\partial u}{\partial t} +Q_1u^2 +C_1u^3 +C_2 u^2 u_t +C_3 u_t^3,
\label{eq:3.18}
\end{equation}
where $\mu$ and $\nu$ are the control parameters, $b>1$, $Q_1, C_1, C_2$ and $C_3$ are constants coefficients with $C_1<0$ to make $\mathcal{F}$ bounded below, and we choose $C_2<0$ and $C_3<0$ to make $\mathcal{F}$ decrease with time at large values of $u$ or $u_t$. 

Now that we have identified conditions on the coefficients of the nonlinear terms in order to ensure global stability, we look at our second goal: to access all the possible dynamical behaviours close to a TB bifurcation. The process to check this in detail is discussed in section \ref{sec:Relate the model to  DK cases}. In order to increase the number of scenarios that are possible, we add one quadratic nonlinear term and two cubic nonlinear terms as shown below. 
 \begin{equation}\label{eq:3.19}
	\frac{\partial^2 u}{\partial t^2} = \mathcal{M}_1 u + \mathcal{M}_2 u_t+Q_1u^2 + C_1 u^3+C_2 u^2 u_t +C_3 u_t^3 +Q_2 u u_{xx}  + C_4(u_x)^2 u_t  + C_5u u_x u_{tx},
	\end{equation}
We do not calculate the Lyapunov functional for this updated model, but rather rely on the conditions that we have derived previously in terms of $C_1$, $C_2$ and $C_3$ to provide global stability. 

\section{Reduction to the Takens--Bogdanov normal form}
\label{sec:TB}
In this section, we use weakly nonlinear theory to reduce our model PDE from Eqn.\,(\ref{eq:3.19}) to the TB normal form in Eqn.\,(\ref{eq:1.1}). We consider $\epsilon\ll1$ to be a small parameter that parameterises the small amplitude solution such that $u = \mathcal{O}(\epsilon)$. This solution can be found in the neighbourhood of the TB bifurcation point where $\mu$ and $\nu$ vary with values of the order of $\mathcal{O}(\epsilon^2)$. This scaling satisfies both the normal forms of a pitchfork and a Hopf bifurcation and has been used to analyse the TB problem previously \citep{knobloch1981nonlinear,dangelmayr1987takens}. Therefore, we scale the field, time and the parameters $\mu$ and $\nu$ as follows:
\begin{equation}\label{eq:4.1}
u = \epsilon u_1 + \epsilon^2 u_2 + \epsilon^3 u_3 + \epsilon^4 u_4 +\cdots, \quad \frac{\partial}{\partial t} \rightarrow \epsilon \frac{\partial}{\partial t}, \quad  \mu \rightarrow \epsilon^2 \mu_2,  \quad \nu \rightarrow \epsilon^2 \nu_2.
\end{equation}
By substituting the scalings into the governing equation (\ref{eq:3.19}), we get the following equation, where we write the terms up to $\mathcal{O}(\epsilon^4)$ explicitly. 
\begin{equation}
	\begin{aligned}
		\frac{\partial^2  \left( \epsilon^3 u_1 + \epsilon^4 u_2 \right) }{\partial t^2}&=
		\mu_2\left( \epsilon^3 u_1 + \epsilon^4 u_2   \right ) 
		- \left(1+\frac{\partial^2}{\partial x^2}\right)^2 \left(\epsilon u_1 + \epsilon^2 u_2 + \epsilon^3 u_3 \right) \\
		&\qquad{}+ \nu_2 \left(\epsilon^4 \frac{\partial u_1}{\partial t} \right) 
		-  b\left(1+\frac{\partial^2}{\partial x^2}\right)^2\left(\epsilon^2 \frac{\partial u_1}{\partial t} + \epsilon^3 \frac{\partial u_2}{\partial t} + \epsilon^4 \frac{\partial u_3}{\partial t} \right) \\
		& \qquad{}+ Q_1\left(\epsilon^2 u_1^2 + 2\epsilon^3 u_1 u_2 + \epsilon^4 u_2^2 + 2 \epsilon^4 u_1 u_3 \right)  \\
	&\qquad{}+Q_2 \left(\epsilon^2 u_1 \frac{\partial^2 u_1}{\partial x^2} +\epsilon^3 \left(u_1  \frac{\partial^2u_2}{\partial x^2}  + u_2  \frac{\partial^2 u_1}{\partial x^2} \right) +\epsilon^4 \left(u_1\frac{\partial^2 u_3}{\partial x^2} +u_2\frac{\partial^2 u_2}{\partial x^2}+u_3\frac{\partial^2 u_1}{\partial x^2}\right)  \right)  
\\
		& \qquad{}+ C_1 \left(\epsilon^3 u_1^3 + 3\epsilon^4 u_1^2 u_2 \right)  
		 +C_2\left(\epsilon^4 u_1^2 \frac{\partial u_1}{\partial t}  \right )  
		+C_4 \left(\epsilon^4 u_1  \frac{\partial u_1}{\partial t} \frac{\partial^2 u_1}{\partial x^2} \right) \\ 
		&\qquad{}+C_5 \left(\epsilon^4 u_1  \frac{\partial u_1}{\partial x} \frac{\partial^2 u_1}{\partial x \partial t} \right) + \mathcal{O}(\epsilon^5)\,.
		\label{eq:4.100}
	\end{aligned}
	\end{equation}
Note that $C_3$ is not present in the above equation since it contributes only at the sixth order of $\epsilon$. Matching terms of order $\epsilon$ and $\epsilon^2$ we get
\begin{align}
\mathcal{O}(\epsilon): \ \ 0 &=\mathcal{L} u_1, \label{e:4.4}\\
\mathcal{O}(\epsilon^2):\ \ 	0 &= \mathcal{L} u_2 + b\mathcal{L} \frac{\partial u_1}{\partial t} + Q_1 u_1^2+Q_2  u_1 \frac{\partial^2 u_1}{\partial x^2}, \label{eq:4.11}
	\end{align}
where  $\mathcal{L}$ is the linear operator defined as
\begin{equation} \label{eq:4.5}
\mathcal{L}= -\left(1+\frac{\partial^2}{\partial x^2}\right)^2\,.
\end{equation}
The linear Eqn.\,(\ref{e:4.4}) obtained at $\mathcal{O}(\epsilon)$ can be solved for $u_1$ by assuming exponential solutions of the form 
\begin{equation}
u_1(x,t) = F_1(t) e^{ix} +\bar{F}_1(t) e^{-ix}\,.  \label{eq:4.7}
\end{equation}
Here $F_1$ and its complex conjugate are functions of time. Substituting this and its derivatives into Eqn.\,(\ref{eq:4.11}), we can solve for $u_2$ as 
\begin{equation}
u_2(x,t) = \frac {\left(Q_1-Q_2\right)}{9} F_1^2e^{2ix}+ G_1 e^{ix}+  2 (Q_1-Q_2) |F_1|^2 + \bar{G_1} e^{-ix}+\frac{\left(Q_1-Q_2\right)}{9}\bar{F_1}^2 e^{-2ix}, \label{eq:4.13}
\end{equation}
where $G_1$ and $\bar{G}_1$ are functions of time. 
 
At $\mathcal{O}(\epsilon^3),$ we have 
\begin{equation} \label{eq:4.14}
\frac{\partial^2 u_1}{\partial t^2}= \mu_2 u_1+\mathcal{L}  u_3 +b\mathcal{L} \frac{\partial u_2}{\partial t} + 2 Q_1 u_1 u_2 
+Q_2\left(u_1  \frac{\partial^2u_2}{\partial x^2}  + u_2  \frac{\partial^2 u_1}{\partial x^2}\right)+ C_1 u_1^3.
\end{equation}
We can solve for $u_3$ by assuming it to be of the form 
\begin{align} \label{eq:4.15}
u_3(x,t)& = H_3(t)e^{3ix}+ H_2(t) e^{2ix} + H_1(t) e^{ix} +H_0(t)+ \bar{H_1}(t) e^{-ix} +\bar{H_2}(t) e^{-2ix} +\bar{H_3}(t) e^{-3ix},
\end{align}
The equation (\ref{eq:4.14}) has contributions into the $e^{ix}$ lengthscale. However, this lengthscale has already been accounted for in the solution for $u_1$. Therefore, we need to enforce the condition that the coefficient of $e^{ix}$ terms are zero as a solvability condition. Substituting (\ref{eq:4.7}), (\ref{eq:4.13}) and (\ref{eq:4.15}) into (\ref{eq:4.14}) and collecting the terms that contribute at the $e^{ix}$ lengthscale, we get 
\begin{equation} \label{eq:4.10}
\frac{\partial^2 F_1}{\partial t^2} = \mu_2 F_1 + A |F_1|^2 F_1,
\end{equation}
where 
\begin{equation} \label{eq:4.16}
A=\left(Q_1-Q_2\right) \left(\frac{38}{9}Q_1- \frac{23}{9}Q_2\right)+3C_1. 
\end{equation}
We can then solve for the unknowns $H_{0,2,3}$ by collecting the coefficients of the constant, $e^{2ix}$ and $e^{3ix}$ respectively as
 \begin{equation} \label{eq:4.19}
H_0= 2\left(Q_1-Q_2\right) \left(F_1\bar{G_1}+ \bar{F_1} G_1 -b\left(\frac{\partial F_1}{\partial t} \bar{F_1} + F_1 \frac{\partial \bar{F_1}}{\partial t}\right)\right).
 \end{equation}
 \begin{equation} \label{eq:4.18}
H_2= \frac{2\left(Q_1-Q_2\right)}{9} \left(F_1G_1-bF_1\frac{\partial F_1}{\partial t}\right),
\end{equation}	
 \begin{equation} \label{eq:4.17}
	H_3= \frac{1}{64}\left(\frac{2}{9}Q_1\left(Q_1-Q_2\right)-\frac{5}{9}Q_2\left(Q_1-Q_2\right)+C_1\right)F_1^3.
		\end{equation}

 At  $\mathcal{O}(\epsilon^4),$ we have 
 \begin{equation}
 \begin{aligned}
 \frac{\partial^2 u_2}{\partial t^2}&= \mu_2 u_2 +\mathcal{L}  u_4+\nu_2 \frac{\partial u_1}{\partial t} + b\mathcal{L} \frac{\partial u_3}{\partial t} + Q_1 \left( u_2^2 + 2 u_1 u_3\right)
  \\&\qquad{}+Q_2 \left(u_1\frac{\partial^2 u_3}{\partial x^2} +u_2\frac{\partial^2 u_2}{\partial x^2}+u_3\frac{\partial^2 u_1}{\partial x^2}\right)  +3 C_1 u_1^2 u_2+C_2 u_1^2 \frac{\partial u_1}{\partial t} \\&
 \qquad{}+C_4  u_1  \frac{\partial u_1}{\partial t} \frac{\partial^2 u_1}{\partial x^2}
+C_5  u_1  \frac{\partial u_1}{\partial x} \frac{\partial^2 u_1}{\partial x \partial t}.
 \label{eq:4.20}
 \end{aligned}
 \end{equation}
The solvability condition from this order of equations is the $e^{ix}$ component of (\ref{eq:4.20}) as below.
 \begin{equation} \label{eq:4.21}
 \begin{aligned}
 \frac{\partial^2 G_1}{\partial t^2}& =\mu_2 G_1+ \nu_2 \frac{\partial F_1}{\partial t}+N F_1^2 \bar{G_1} + P |F_1|^2G_1
 + C\left(\frac{\partial F_1}{\partial t}\bar{F_1} + F_1 \frac{\partial \bar{F_1}}{\partial t}\right) F_1 + D |F_1|^2 \frac{\partial F_1}{\partial t},
 \end{aligned}
 \end{equation}
 where 
 \begin{align} 
 N&= \left(Q_1-Q_2\right) \left(\frac{38}{9}Q_1- \frac{23}{9}Q_2\right)+3C_1, \label{eq:4.22} \\
  P&= \left(Q_1-Q_2\right) \left(\frac{76}{9}Q_1- \frac{62}{9}Q_2\right)+6C_1,\label{eq:4.23} \\
  C&= 2b\left(Q_1-Q_2\right)\left(-2Q_1+Q_2\right)+C_2-C_4+C_5,   \label{eq:4.24}\\
   D&= b\left(Q_1-Q_2\right)\left(\frac{-4}{9}Q_1+ \frac{10}{9}Q_2\right)+C_2+3C_4-C_5.  \label{eq:4.25} 
   \end{align}
 In order to ensure that both the solvability conditions arising from both the $\mathcal{O}(\epsilon^3)$ and $\mathcal{O}(\epsilon^4)$ equations are satisfied, we use a reconstitution procedure \citep{rucklidge2009design} to combine equations (\ref{eq:4.10}) and (\ref{eq:4.21}) into a single PDE by defining a new variable
\begin{equation} \label{eq:4.26}
z = \epsilon F_1 + \epsilon^2 G_1.
\end{equation}
By unscaling time and  the parameters according to 
\begin{equation}  \label{eq:4.27}
  \frac{\partial }{\partial t}\rightarrow \frac{1}{\epsilon} \frac{\partial }{\partial t} \qquad   
\mu_2 \rightarrow  \frac{1}{\epsilon^2} \mu \qquad \text{ and} \ \nu_2 \rightarrow  \frac{1}{\epsilon^2} \nu. 
\end{equation}
We get
\begin{equation}  \label{eq:4.28}
\begin{aligned}
\frac{\partial^2 z}{\partial t^2}&= \mu z +\epsilon \nu \frac{\partial F_1}{\partial t} + \epsilon^3 A |F_1|^2F_1 + \epsilon^4 N F_1^2 \bar{G_1} + \epsilon^4 P |F_1|^2G_1 
\\&\qquad +\epsilon^3C\left(\frac{\partial F_1}{\partial t} \bar{F_1}+ F_1\frac{\partial \bar{F_1}}{\partial t}\right)F_1 + \epsilon^3D|F_1|^2\frac{\partial F_1}{\partial t}. 
\end{aligned}
\end{equation}
 Substituting  $F_1 = \frac{z}{\epsilon} - \epsilon G_1$ in (\ref{eq:4.28}) and collecting terms of $\mathcal{O}(1)$ in the above equation, we get the TB normal form as
 \begin{equation}
 \frac{\partial^2 z}{\partial t^2}  = \mu z + \nu \frac{\partial z}{\partial t} +  A |z|^2 z +    C \left( \frac{\partial z}{\partial t} \bar{z}+ z  \frac{\partial \bar{z}}{\partial t}\right) z +   D |z|^2  \frac{\partial z}{\partial t}, 
  \label{eq:4.29}
 \end{equation}
 where $A,C$ and $D$ are given by (\ref{eq:4.16}), (\ref{eq:4.24}) and (\ref{eq:4.25}), respectively. 
 
\section{Accessing possible different dynamical behaviour near the TB bifurcation} \label{sec:Relate the model to  DK cases}

In this section, we identify parameter combinations in the model at which we can observe different dynamical behaviours close to the TB bifurcation. In order to do this, we relate the parameters in the current model with those investigated in detail by \citet{dangelmayr1987takens}. In the rest of this paper, we refer to this paper as DK. DK identify different bifurcation scenarios close to a TB bifurcation and classify them in terms of the value of coefficient $A$ as well as the ratio $D/(2C+D)$ where $2C+D$ is defined as $M$. The ratio can be related to the coefficients in our model as below. 
\begin{equation}
\frac{D}{M} =\frac{  {-4} bQ_1^2+ {14} bQ_1 Q_2- {10} b Q_2^2 + 9 C_2+ 27 C_4- 9 C_5 }{ {-76} bQ_1^2+ {122} bQ_1 Q_2- {46} b Q_2^2 + 27 C_2+ 9 C_4+ 9 C_5}.
\label{eq:3.30}
\end{equation}

We consider a few special cases below to illustrate the need for certain nonlinear terms in the model. First, let us consider the case with $Q_2=C_4 = C_5=0$. Then using expressions in Eqns.\,(\ref{eq:4.16}) and (\ref{eq:4.25}), the values of $A$ and $D$ will be negative and the fraction ${D}/{M}$ is always bounded between the values
$$ \frac{1}{19} < \frac{D}{M} < \frac{1}{3}. $$
In this range of ratios, we can only access one type of bifurcation behaviour near the TB bifurcation identified from the classification given in DK as $A<0$, case \RN{2}${-}$.

\begin{figure}[t!]
  	\centering 
  	\includegraphics[width=0.9\linewidth]{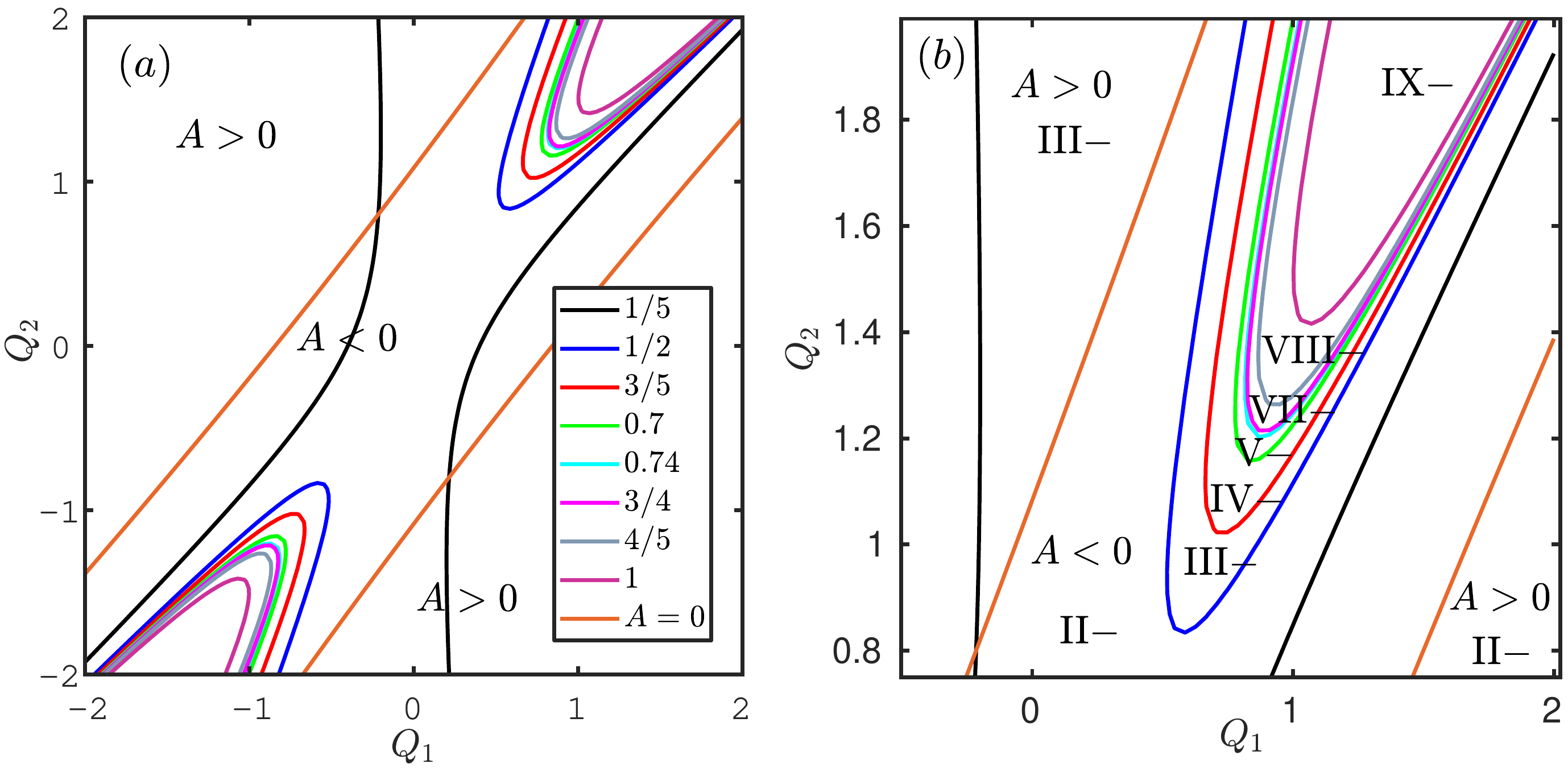}
  	\caption{($a$) Plot of the contours of $A=0$ in orange from (\ref{eq:4.16}) and the different values of $\frac{D}{M}= c$ where $c= \frac{1}{5},\frac{1}{2},\frac{3}{5},0.7,0.74,\frac{3}{4},\frac{4}{5}$ from (\ref{eq:4.1-2-2}) where $C_1=C_2=C_3=-1$ and $b=2$. The regions between each curve correspond to the enumerated \RN{2}$-$,...\RN{9}$-$ cases in DK. ($b$) Zoomed in view of the top right corner of ($a$).}
	\label{fig:4.2}
  	\end{figure}

Secondly, we consider the case with only $C_4 =C_5=0$, which gives the expression for the ratio $D/M$ as
  \begin{equation} \label{eq:4.1-2-2}
\frac{D}{M} =\frac{  {-4} bQ_1^2+ {14} bQ_1 Q_2- {10} b Q_2^2 + 9 C_2 }{ {-76} bQ_1^2+ {122} bQ_1 Q_2- {46} b Q_2^2 + 27 C_2}\,,
  \end{equation}	
Considering that Lyapunov stability requires $C_1,C_2$ and $C_3$ to be negative, we consider the case with  $C_1=C_2=C_3=-1$ and $b=2$. By plotting contours of $A$ and $\frac{D}{M}=c$, where $c=\frac{1}{5}, \frac{1}{2}, 0.7,0.74, \frac{3}{4}, \frac{4}{5},1$, we can obtain a range of normal form cases. Figure \ref{fig:4.2} shows the regions of different cases accessible in this case as a function of the two quadratic coefficients $Q_1$ and $Q_2$. From this figure, we see that we can access all bifurcation scenarios with $A<0$, while in the case with $A>0$, we are unable to access cases \RN{1}$-$ and \RN{4}$-$. For other choices of parameters, it is possible to access all cases on the normal form with $M<0$, while still satisfying the Lyapunov stability requirement. However, cases of the normal form with $M>0$ still remain inaccessible with $C_4=C_5=0$.

Thirdly, when we allow for all nonlinear terms in the model, we can access all cases of the normal form listed in DK, i.e., 18 cases with $A<0$ and 8 cases with $A>0$ both with $M<0$ and $M>0$. As a first reference, we give a list of parameters that allow us to reach all the cases with $M<0$ in Table~\hbox{\ref{tab:con-1}}. Each of these cases has a different bifurcation scenario close to the TB bifurcation. 

Of all these cases, we choose two to look at in detail. Our choice is guided by the predicted stability diagrams from DK, which indicate the possibility of interesting coexistence regions of different states, e.g., coexistence of the trivial and a patterned steady state (SS), coexistence of SS and standing wave (SW), etc. Cases of particular interest are $\RN{4$-$}$ with $A>0$ and $\RN{1$-$}$  with $A<0$ in DK and marked in blue in Table \ref{tab:con-1}.

\begin{table}[t] 
\centering
\begin{tabular}{ @{} c c c c c c c c@{}} 
\toprule
Case & \multicolumn{7}{c}{Coefficients of the nonlinearities}   \\ [0.5ex]
 in  DK     & $Q_1$ & $Q_2$ & $C_1$&$C_2$&$C_3$&$C_4$&$C_5$\\ [0.5ex]
 \toprule 
 $A>0$ & & & & & & &   \\
 \vspace{0.1cm}
\RN{1}$-$&$0.8$ & $-0.5$ & $-1$&$-0.1$&$-1$&$-0.1$&$-5$ \\
\vspace{0.1cm}
 \RN{2}$-$&$1.5$ & $0.5$ & $-1$&$-1$&$-1$&$0$&$0$ \\
\vspace{0.1cm}
 \RN{3}$-$& $0.1$ & $1.5$ & $-1$&$-1$&$-1$&$0$&$0$\\
\vspace{0.1cm}
 {\color{blue}\RN{4}$-$}& {\color{blue}$0.9$}& {\color{blue}$-0.2$}& {\color{blue}$-0.2$}&{\color{blue}$-1$}&{\color{blue}$-1$}&{\color{blue}$-0.5$}&{\color{blue}$6$}\\
 \toprule 
 \vspace{0.1cm}
  $A<0$ & & & & & & &  \\
\vspace{0.1cm}
{\color{blue} \RN{1}$-$}&  {\color{blue}$0.8$}& {\color{blue}$0.5$}& {\color{blue}$-1$}&{\color{blue}$-0.1$}&{\color{blue}$-1$}&{\color{blue}$-0.1$}&{\color{blue}$-5$} \\
\vspace{0.1cm}
\RN{2}$-$& $0.5$ & $0$ &$-1$&$-1$&$-1$&$0$&$0$ \\
\vspace{0.1cm}
 \RN{3}$-$& $0.6$ & $0.9$&$-1$&$-1$&$-1$&$0$&$0$ \\
\vspace{0.1cm}
\RN{4}$-$& $0.8$ & $1.1$ &$-1$&$-1$&$-1$&$0$&$0$\\
\vspace{0.1cm}
 \RN{5}$-$& $0.85$ &$1.2$ &$-1$&$-1$&$-1$&$0$&$0$\\
\vspace{0.1cm}
 \RN{6}$-$& $0.9$ &$1.21$ &$-1$&$-1$&$-1$&$0$&$0$\\
\vspace{0.1cm}
 \RN{7}$-$& $0.9$ &$1.25$ &$-1$&$-1$&$-1$&$0$&$0$ \\
\vspace{0.1cm}
 \RN{8}$-$& $1$ & $1.4$ &$-1$&$-1$&$-1$&$0$&$0$\\
\vspace{0.1cm}
 \RN{9}$-$&$1.1$ & $1.5$ &$-1$&$-1$&$-1$&$0$&$0$ \\
\bottomrule 
\end{tabular}
\caption{Examples of parameter values in the model (\ref{eq:3.19}) to access different cases in the normal form in DK. Only cases with $M<0$ are listed here. The instances in blue indicate the cases we considered in this paper with the main result.}
\label{tab:con-1}
\end{table} 


\section{Localized SS} 
\label{sec:lss}

\begin{figure}[t!]
  	\centering 
  	\includegraphics[width=0.95\linewidth]{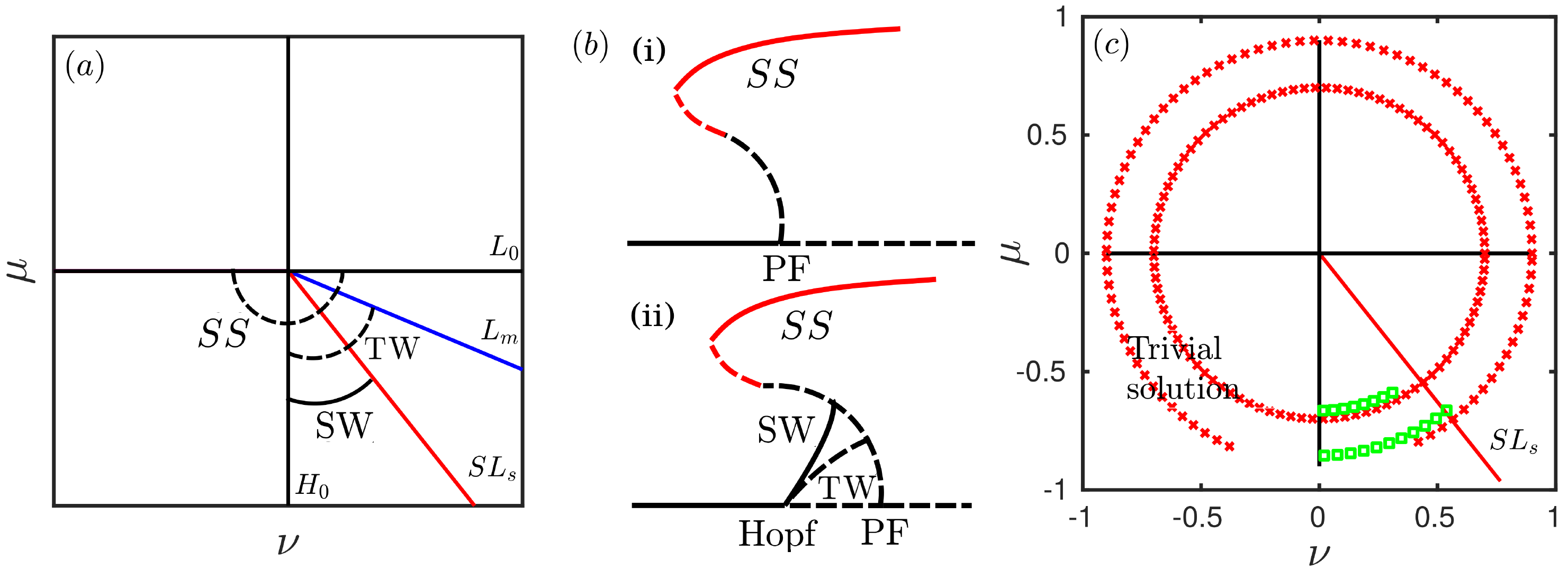}
%
    \caption{(a) Sketch of the stability diagram for case \RN{4$-$} with $A>0$ in $(\nu,\mu)$-plane and (b) the corresponding bifurcation diagram from \citep{dangelmayr1987takens}, where the panel (i) represents the bifurcation above the diagonal in the $(\nu,\mu)$-plane and the panel (ii) represents the bifurcation below the diagonal in the $(\nu,\mu)$-plane. In this bifurcation diagrams, the black lines (solid for stable solutions, dashed for unstable solutions) are from DK, while the red lines are inferred from our system given Lyapunov stability. (c) Plot of the solution from solving the model (\ref{eq:3.19}) by time stepping with parameters values $Q_1=0.9, Q_2=-0.2,C_1=-0.2, C_2=C_3=-1, C_4=-0.5,C_5=6$ and $b=2$ for radius $r=0.7,0.9$. A Hopf bifurcation occurs at $\theta=270^{\circ}$ and a pitchfork bifurcation occurs at $\theta=180^{\circ}$ and $\theta=0^{\circ}$. The  red x and green square refer to extended SS and SW solutions, respectively. The half line $SL_s$ is the line of heteroclinic bifurcations where SW joins the small-amplitude unstable SS.} 
    \label{fig:5.1}
\end{figure}

The first case that we discuss is the one labelled as case $\RN{4$-$}$ with $A>0$ in DK. Figure \ref{fig:5.1}($a$) shows the stability in $(\nu,\mu)$-plane as predicted by the TB normal form in DK. Here we observe that there is a small unstable branch of SS that lies in the third quadrant (where the trivial state is stable). Further, we observe a stable branch of SW in the fourth quadrant along with the unstable branch of SS. We identify this bifurcation scenario to be interesting as we expect the possibility for localised steady states in a background of trivial state (LSS-TS) in the third quadrant and the possibility for localised steady state in a background of standing waves (LSS-SW) in the fourth quadrant. 

Figure \ref{fig:5.1}($b$) shows the predicted bifurcation diagrams (in black lines) from the TB normal form equation. The two cases correspond to the cases of variation of parameters along the diagonal above and below the line $\mu=\nu$ respectively. The bifurcation above the diagonal in $(\nu,\mu)$-plane (see Figure \hbox{\ref{fig:5.1}(b)(i)} ) has only a subcritical SS branch from a pitchfork bifurcation at $\mu=0$. The trivial state is stable in the region where $\mu<0$ and $\nu<0$. The bifurcation below the diagonal in $(\nu,\mu)$-plane (see Fig.\,\ref{fig:5.1} b, right panel) has a subcritical SS branch from a pitchfork bifurcation at $\mu=0$. Stable SW and unstable TW bifurcate from the trivial state at a Hopf bifurcation where $\nu=0, \mu<0$. The stable SW branch  terminates on the subcritical SS branch with the formation of a heteroclinic orbit at a global bifurcation $SL_s$ connecting two saddles (the names of these bifurcations follows DK). The unstable TW  branch terminates at the subcritical SS at $L_m$. This scenario has been investigated analytically and numerically in thermosolutal convection~ \citep{knobloch1981nonlinear, huppert1976nonlinear, da1981oscillations, knobloch1986transitions} and is important because it was an early example of the discovery and analysis of how a Shil'nikov heteroclinic orbit \citep{knobloch1986transitions} can lead to chaotic dynamics, though this is in a different parameter regime from that which we will investigate. To the bifurcation diagrams in Figure \hbox{\ref{fig:5.1}($b$)}, we add predictions (from global stability requirements) of a large amplitude stable branch of steady patterned state (in red lines). This figure now illustrates the possible coexistence regions that could allow for the localised states detailed in the previous paragraph. 

In order to confirm the existence of a stable large amplitude branch of patterned state (SS), we first look at asymptotic states accessible via time stepping. We treat the model numerically by discretizing the PDE both in time and space.  In space, we discretize the model using spectral methods \citep{cox2002exponential, kassam2005fourth, canuto2012spectral, hussaini1987spectral} and fast Fourier transform (FFT) with 16 grid points per wavelength. In time, we discretize the model using the exponential time differencing method (ETD)  \citep{cox2002exponential, kassam2005fourth, canuto2012spectral, hussaini1987spectral}. The ETD method solves the linear parts of the PDEs exactly followed by a second order approximation of the nonlinear parts. We  compare the stability region from solving the model with the stability region obtained from the normal form \citep{dangelmayr1987takens}. To do this we solve the PDE and plot the type of solutions in $(\nu,\mu)$-plane, using polar coordinates defined as
\begin{equation} \label{eq:munu}
\nu= r \cos(\theta), \qquad \qquad \mu=r \sin(\theta),
\end{equation}
where $r$ is the radius that controls how far $\nu$ and $\mu$ are from the TB point, and $\theta$ is the angle that controls  the position of $\nu$ and $\mu$ in the $(\nu,\mu)$-plane.  Note that the Hopf bifurcation occurs at $\nu=0$ and $\mu<0$ which correspond to $\theta=270^{\circ}$. The pitchfork bifurcation occurs at $\mu=0$ which correspond to $\theta=0^{\circ}$ and $\theta=180^{\circ}$. 

Figure \ref{fig:5.1}($c$) shows the results of starting time stepping from different initial conditions for a variety of system parameters at two different radii $r=0.7$ and $r=0.9$. At radius $r=0.7$, we start from initial conditions close to a pitchfork bifurcation at $\theta=180^{\circ}$. We are able to obtain large amplitude SS solutions (shown as red crosses) as well as recover the trivial state (not shown with markers). The trivial state is stable when $\mu<0$ and $\nu<0$ until we reach a Hopf bifurcation close to $\theta=270^{\circ}$. At this bifurcation, the trivial state loses stability and a new branch of SW are created (shown as green crosses). The amplitude of the SW branch increases with increasing $270^{\circ}<\theta<297^{\circ}$ which is close to the prediction of an $SL_s$ bifurcation at an angle of $\theta=308^{\circ}$ from the normal form analysis. The complete circle of red crosses observed for the large amplitude SS in Fig.\,\ref{fig:5.1}($c$) at $r=0.7$ indicates that the solution branch of large amplitude SS is disconnected from the low amplitude SS solution branch when viewed as a function of $\theta$. 

At a slightly larger radius with $r=0.9$, time stepping identifies similar coexistence of trivial state and large amplitude SS solutions in the third quadrant. However, the large amplitude SS solutions exist only till $\theta=245^{\circ}$. At $\theta=270^{\circ}$, we encounter the Hopf bifurcation, as before, resulting in the branch of SW solutions. The branch of SW exists in the range of $270^{\circ}<\theta<309^{\circ}$ and terminate close to the prediction of the $SL_{s}$ bifurcation. We find that we are able to recover the large amplitude SS branch again from $\theta=297^{\circ}$. The fact that we are able to identify two $\theta$ values where the large amplitude SS solution branch terminates indicates that they are the locations of saddle-node bifurcations where the large amplitude SS branch connects with the low amplitude SS branch. 


Figure \ref{fig:5.1}($c$) has identified two bistable regions: bistability  between the trivial state and large-amplitude SS when $\mu<0$, $\nu<0$ and bistability between small-amplitude SW and large-amplitude SS in the region between $\nu=0$, $\mu<0$ and the half line $SL_s$. We now look to obtain localized states in these regions. To do this we follow the process. First, we increase the domain size to allow 64 wavelengths 
and perform time stepping to find an extended SS. Then, we use a $\sech$-envelope with different widths to construct several initial conditions and perform time stepping again to obtain nearby dynamically stable localized states. Using this method we are able to get localized steady states, LSS in both of the bistable regions with two different backgrounds. 

\begin{figure}[!t]
  	\centering 
  	\includegraphics[width=0.99\linewidth]{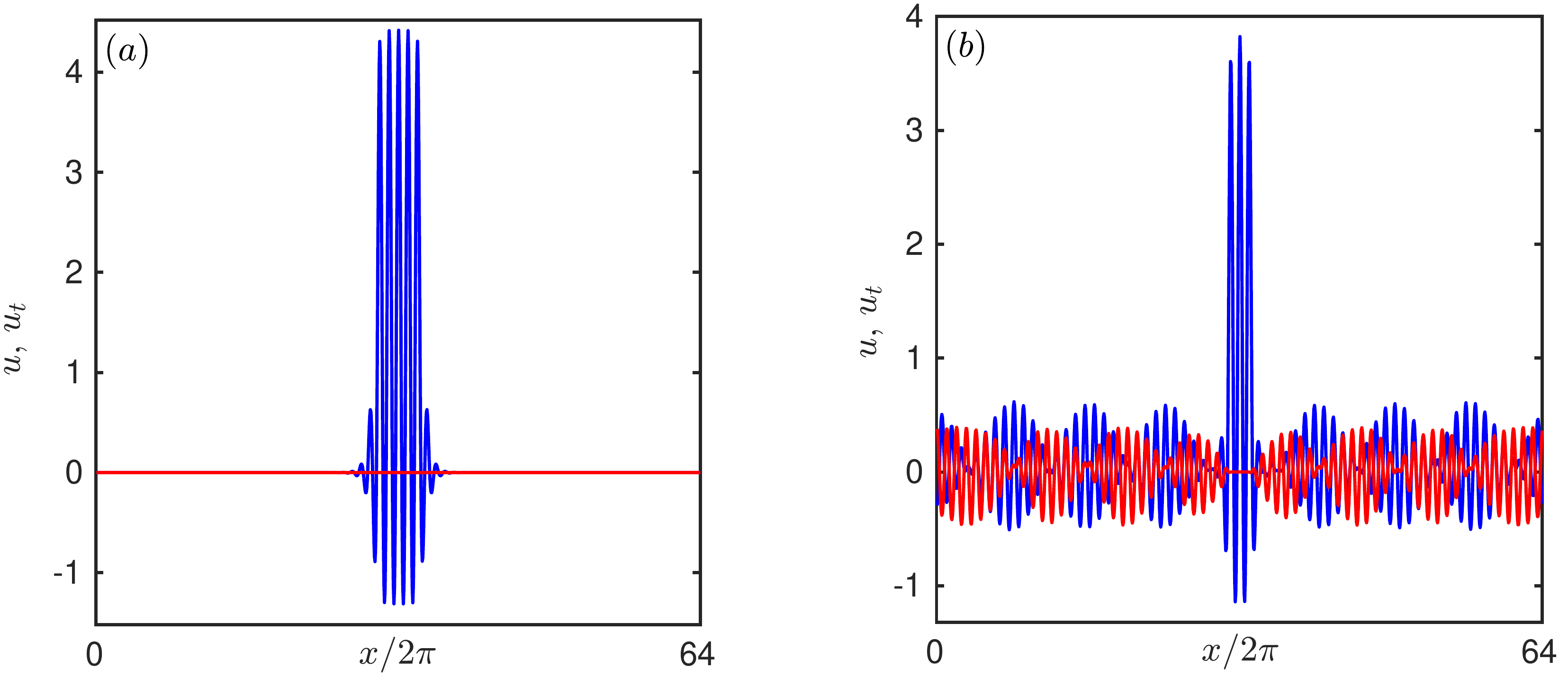}
  \caption{($a$) Localized steady state with trivial solutions background from time stepping at radius 0.7 and $\theta= 220^{\circ}$ where  $\nu= -0.54 , \mu=-0.45$. ($b$) Localized steady state with MW solutions background from time stepping at radius 0.7 and $\theta= 280^{\circ}$ where  $\nu=0.12, \mu= -0.69$. The blue and red curves refer to  $u$ and  $u_t$, respectively. A movie of the state in ($b$) is available at \citep{SuppTBSH}.} 
    \label{fig:5.51}
\end{figure}

First, we find LSS-TS, which has a localised steady state with a trivial state (TS) as a background in the region where the trivial state and large-amplitude SS are both stable ($\mu<0$ and $\nu<0$). Figure \hbox{\ref{fig:5.51} (a)} shows one example of LSS with TS background for radius 0.7 and $\theta=200^{\circ}$ ($\nu= -0.54$, $\mu=-0.45$). There are other  LSS-TS with different widths, depending on the initial conditions, and to investigate these in detail we perform numerical continuation in next section. 

Second, we find LSS with an MW background in the region where the large-amplitude SS and the small-amplitude SW are both stable. The bistability occurs  in the region between the Hopf bifurcation at $\theta= 270^{\circ}$ ($\nu=0, \mu<0$) to the half line $SL_s$ at $\theta\approx 308.4^{\circ}$, as mentioned above.  The small-amplitude MW background solutions  move as a function of time. Initially, the small-amplitude solutions are SW with large-amplitude SS in the middle of the domain. As time increases  the SW turn in to MW. This change occurs because the left-right symmetry of the SW solutions is broken by the SS solutions in the middle. Figure  \ref{fig:5.51} (b) shows one example of LSS with MW background for radius 0.7 and $\theta= 280^{\circ}$ ($\nu=0.12, \mu= -0.69$).  Many widths of this class of solutions can be obtained by altering the initial width of the $\sech$-envelope. 

Unlike the LSS with the trivial state background, in this case two patterns (large-amplitude SS and small-amplitude MW) coexist, which suggests the presence  of a spatial heteroclinic orbit between the SS and MW states. 

\subsection{Numerical continuation} 
In the following we use numerical continuation to compute steady numerical solutions of the model for both the extended SS and the LSS. The method we use is based on Newton iteration and pseudo arclength continuation \citep{Doedel1991}.  
Seting the time-derivative terms to zero in (\ref{eq:3.19}) results in the steady Swift--Hohenberg equation: 
\begin{align} \label{eq:5-7} 
0 = \left(\mu-\left(1+\frac{\partial^2}{\partial x^2}\right)^2\right) u+ Q_1u^2 +Q_2 u u_{xx} + C_1u^3.
\end{align}
We note that even when the solutions depend only on $\mu$, the stability depends on both $\mu$ and $\nu$. 
The initial guesses for the branches of extended SS and LSS are obtained from  time stepping. Since $\theta$ controls the values of $\mu$ and $\nu$ in the $(\nu,\mu)$-plane, we will use $\theta$ as the bifurcation parameter with fixed radius. 
We show the bifurcation diagram of all solutions as functions of  $\theta$ and $\mu$, where $\mu=r \sin(\theta)$, against the norm as a measure of the solutions, where 
\begin{equation} 
||u||_2= \int_{0}^{L} (u(x))^2 dx.
\end{equation}
We will perform continuation at two different radii $r=0.7$ and $r=0.9$ in a one dimensional domain allowing 16 wavelengths with 16 grid points per wavelength. 
\subsubsection{Continuation for radius 0.9}  
\begin{figure}[!t]
  	\centering 
  	\includegraphics[width=0.99\linewidth]{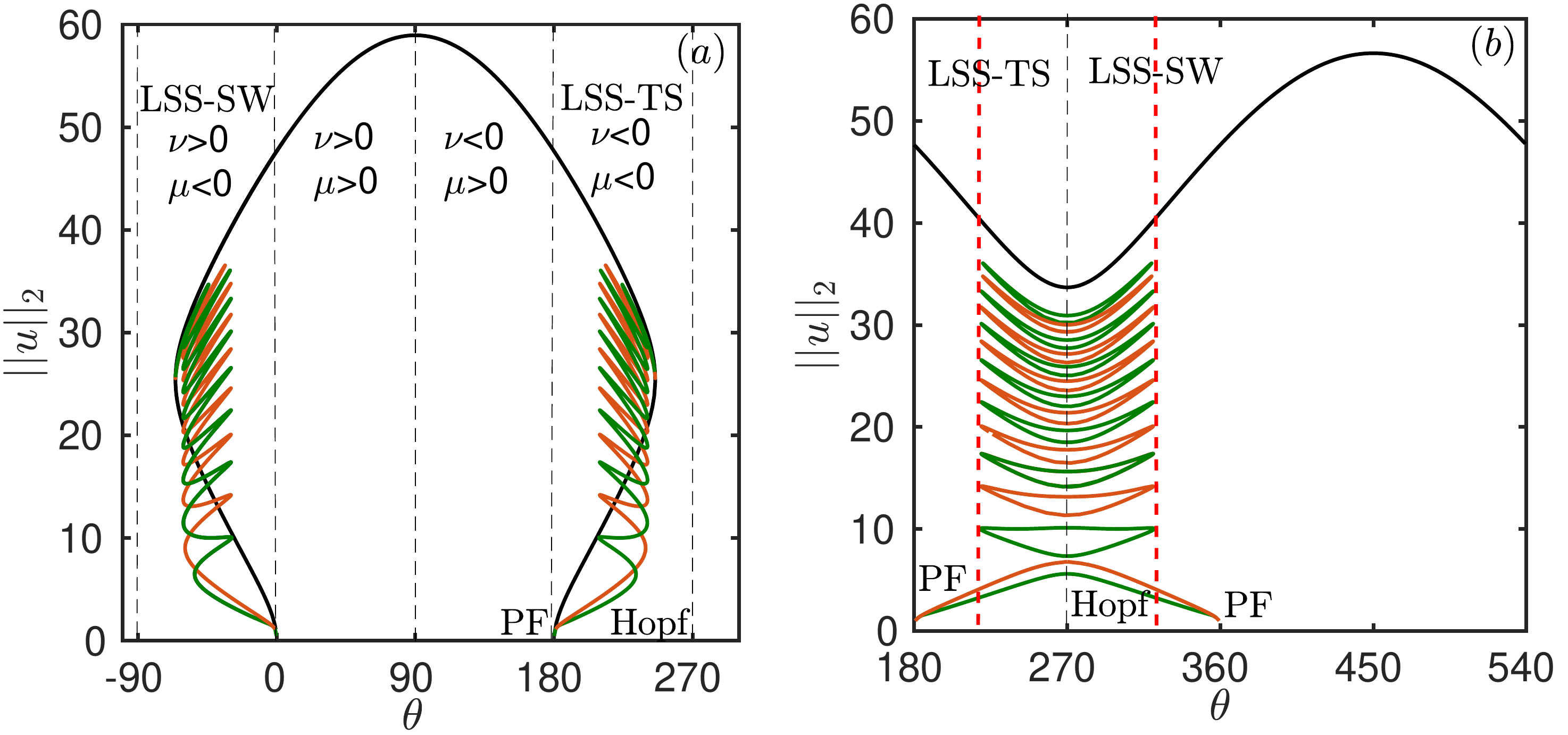}
  \caption{The full branch for extended SS in black and the odd and even  branch in green and brown where $\theta$ is the control parameter and radius (a) 0.9 and (b) 0.7. } 
    \label{fig:5.6}
\end{figure}

\begin{figure}[!t]
	\centering{ 
  	\includegraphics[width=0.97\linewidth]{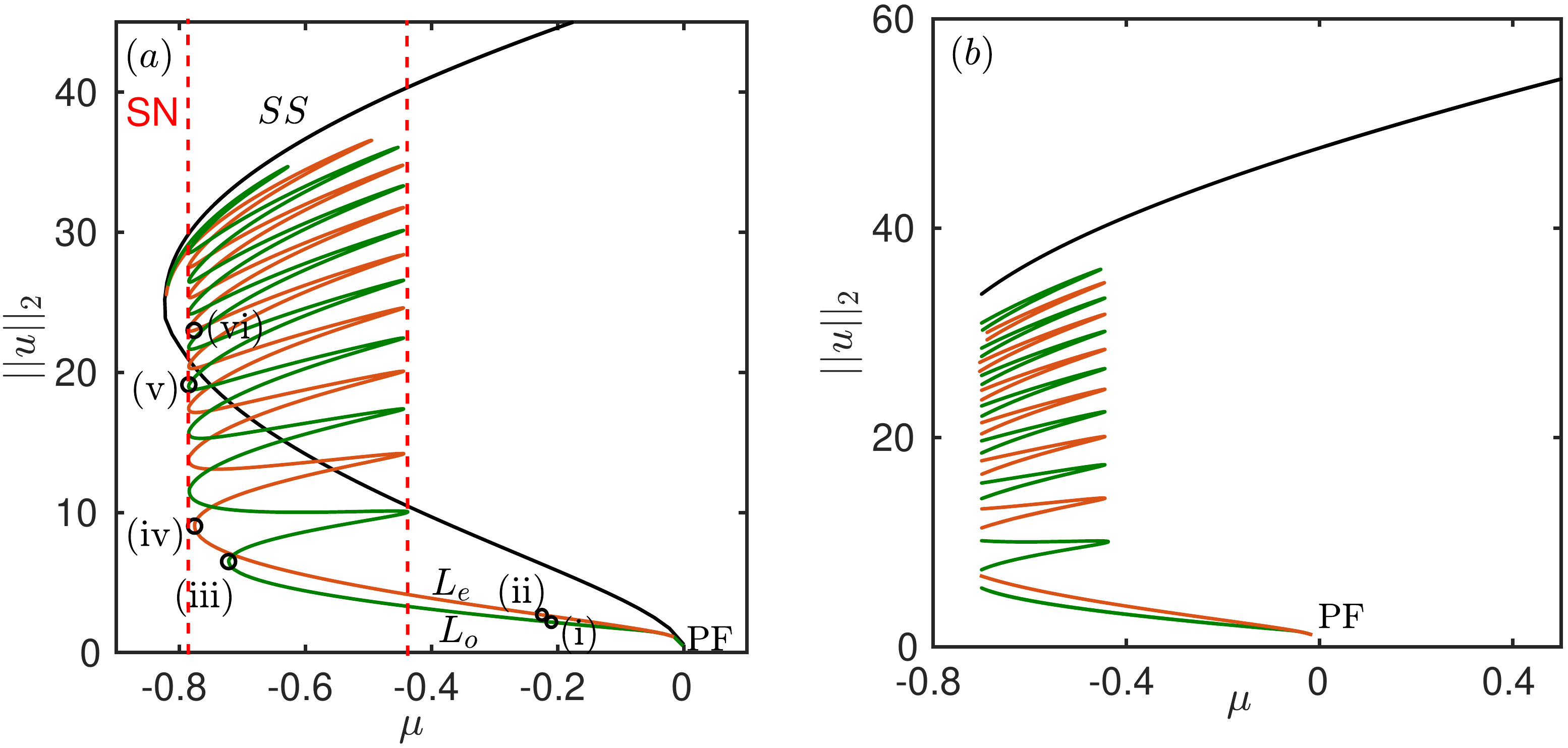}\\
	\vspace{0.4cm}
	\includegraphics[width=0.97\linewidth]{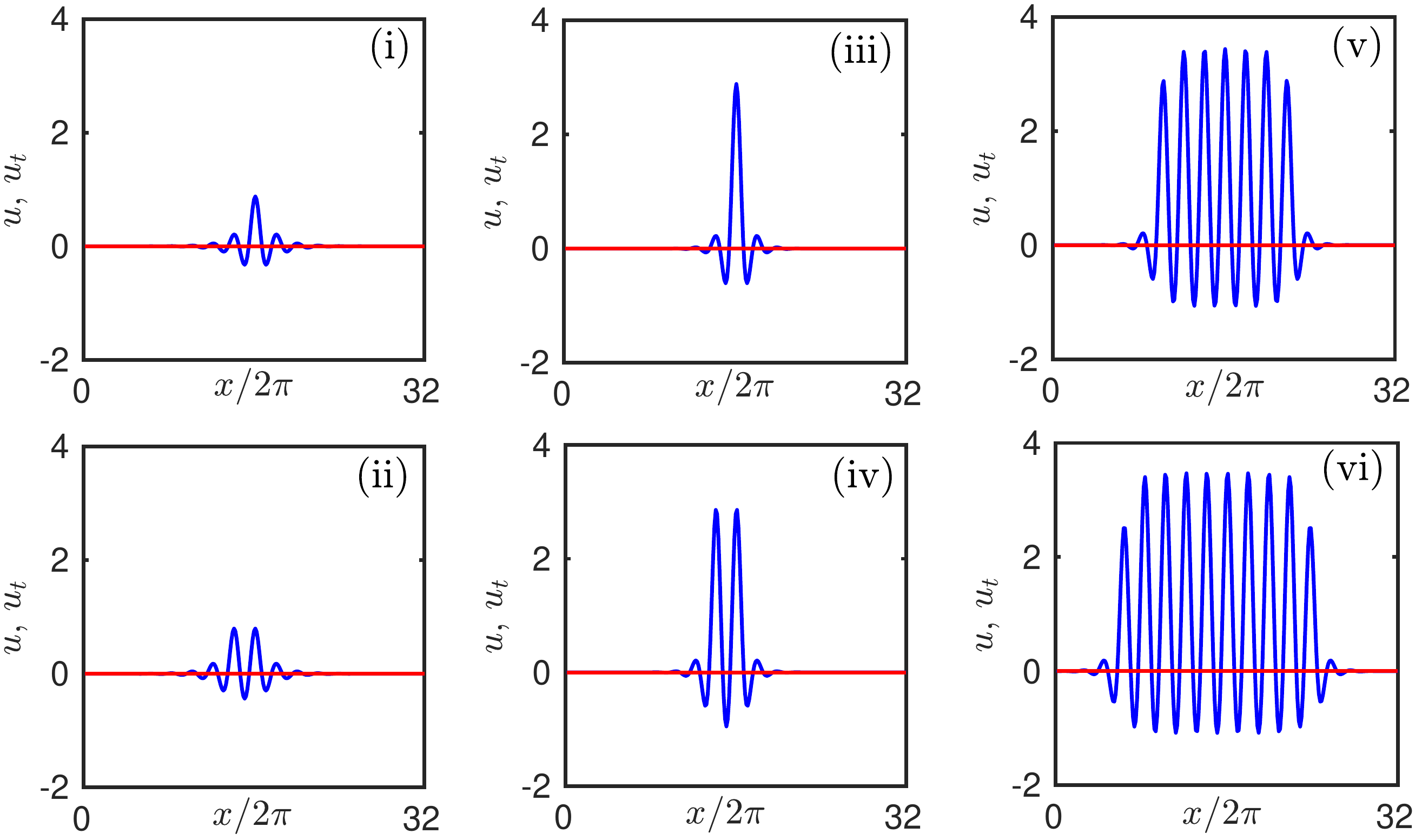}}
%
%
        \caption{The full branch for extended SS in black and the LSS-TS solutions with odd $L_o$ and even $L_e$ branches in green and brown where $\mu$ is the control parameter and radius (i) $r=0.9$ and (ii) $=0.7$. The SS branch bifurcate from trivial state at a pitchfork bifurcation $\mu=0$. The saddle-node point (SN) in ($a$) is $\mu=-0.81$. The Hopf bifurcation occurs at $\mu= -0.9$ where $\nu=0$. (i, iii, v) Show the solutions obtained from the odd branch  $L_o$ with different widths and (ii, iv, vi) show the solutions obtained from the even branch $L_e$. The snaking region in ($a$) is the region between the red dashed lines. }
        \label{fig:5.7}
\end{figure}

Starting from initial guesses obtained from time stepping in numerical continuation with radius $r=0.9$, we perform continuation to obtain the extended SS and the LSS-TS solutions. Figure \ref{fig:5.6} shows the solutions from numerical continuation with $\theta$ as the control parameter in ($a$) for radius $r=0.9$ and ($b$) for radius $r=0.7$. In Fig.\,\ref{fig:5.6}, we represent the full branch of extended SS in black, the LSS branch with even numbers of peaks  $L_e$ in orange and the LSS branch with odd numbers of peaks  $L_o$ in green. Along the odd branch $L_o$ the midpoint $\left(x = {Lx}/{2}\right)$ of the localized state is always a global maximum (with an odd number of maxima), while along
the even branch $L_e$ the midpoint  $\left(x = {Lx}/{2}\right)$ is a global minimum (with an even number of maxima).  Figure \ref{fig:5.7} shows Fig.\,\ref{fig:5.6} (a) for $\mu$ where
 $\mu=d \sin(\theta)$ in more detail with examples of LSS with different widths. 
 
 The extended SS branch emerges subcritically from the trivial state at $\theta =180^{\circ}$ (the pitchfork bifurcation $\mu=0$) and undergoes a saddle-node bifurcation at $\theta= 245.6^{\circ}$ ($\mu=-0.81$). The branch changes at the saddle-node to a large-amplitude stable state. It reaches the maximum amplitude when $\theta=90^{\circ}$ ($\mu=0.9$) and decreases until reaches to the second saddle-node at $\theta= -65.6^{\circ}$ ($\mu=-0.81$). The brach then decrease further until terminates back to a pitchfork bifurcation $\theta=0^{\circ}$  ($\mu=0$).  Because the solutions depend on $\mu$ only there is a symmetry $\theta \rightarrow 180^{\circ} - \theta$.    

Similarly, we use time stepping to obtain the initial guess for the even $L_e$ and odd  $L_o$ branches for localized state and then do the continuation.  Both branches emerge subcritically close to the pitchfork bifurcation with small-amplitude and undergo a series of saddle-node bifurcations producing the homoclinic snaking. Each branch adds an oscillation on each side as it snakes back and forth, until they reach the width of the domain where they terminate on the SS branch, at the saddle-node point.  The snaking region occurs between $\theta=209.5^{\circ}$ ($\mu= -0.44$) and $\theta=240.5^{\circ}$  ($\mu=-0.78$). The Hopf bifurcation occurs when $\theta=270^{\circ}$ (where $\nu=0$ and $\mu=-0.9$). 
Note that, for this choice of radius the homoclinic region occurs away from the Hopf bifurcation. The trivial solution is stable in the region where  $\mu<0$ and $\nu<0$ and unstable anywhere else.
\subsubsection{Continuation for radius 0.7}  
In this section, we will examine the numerical continuation for smaller radius $r=0.7$. The difference here from the previous case that we discussed for $r=0.9$, is that here the saddle-node bifurcation of the SS branch at $\mu=-0.81$ is beyond the minimum value of $\mu$ possible, i.e., $\mu = -0.7$. This implies that the left side of the snaking branch cannot be reached. 

To obtain the numerical branch for LSS as well as the extended SS branch we use the same process as for radius 0.9. We show the results in Fig.\,\ref{fig:5.6} (b) for $\theta$ against the norm. 

Starting from large extended SS as initial guesses in continuation, we find that the branch of extended SS has large-amplitude with maximum at $\theta=90^{\circ}$ ($\mu= 0.7$) and minimum amplitude at $\theta= 270^{\circ}$ ($\mu=-0.7$). This branch continues around the circle from $\theta=0^{\circ}$ to $\theta= 360^{\circ}$ and does not connect to the pitchfork bifurcation, which indicates that there is no saddle-node bifurcation. 

To find the localized branch we start continuation from initial guesses with various numbers of peaks obtained from time stepping. The continuation shows that the  odd and even branches with at least 15 peaks form isolas.  They start from the region where the trivial state stable  ($\mu<0$ and $\nu<0$) and continue to the region where the trivial state not stable after passing the Hopf bifurcation ($\nu=0$ and $\mu<0$). Each isolated branch reach the minimum amplitude at $\theta=270^{\circ}$  ($\mu=-0.7$). The snaking region of isolated branch occurs between  $\theta=219.5^{\circ}$ ($\mu= -0.44$) and $\theta= 320.5^{\circ}$ ($\mu=-0.44$). The two lower branches with one and two peaks bifurcate  close the pitchfork bifurcation  $\theta=180^{\circ}$ ($\mu=0$) with small-amplitude, reach maximum at $\theta = 270^{\circ}$ ($\mu = -0.7$), and then terminate with small-amplitude close to the pitchfork bifurcation  $\theta= 360^{\circ}$ ($\mu=0$).

From time stepping, the LSS with the trivial state background are stable in the region where $\mu<0$ and $ \nu<0$ and become unstable after passing the Hopf bifurcation in the region where $\mu<0$ and $ \nu>0$. After the Hopf bifurcation, LSS-TS exist, but the stable solutions are the LSS with an MW background where there is bistability between SS and SW.  

\begin{figure} [!t]
\centering 
\includegraphics[width=0.5\linewidth]{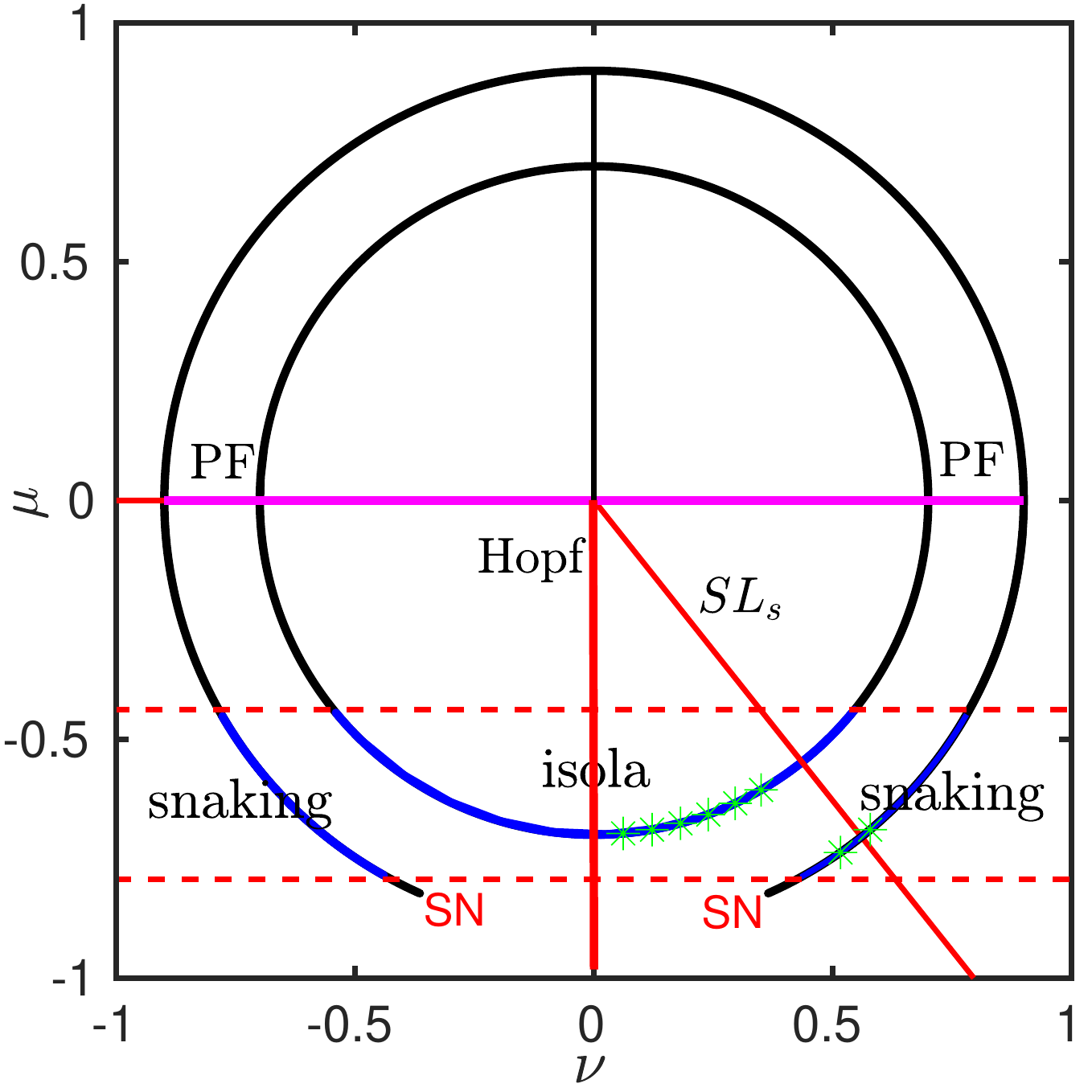}
\caption{$(\nu,\mu)$-plane from solving the PDE using numerical continuation for radii 0.9 and 0.7. The pink and red cures refer to the pitchfork bifurcation and Hopf bifurcation, respectively.  The black curve refers to the extended SS solutions and the blue curve refers to the snaking region for radius $r=0.9$, as shown in Fig.\,\ref{fig:5.7} ($a$)  and the region of isolas for radius $r=0.7$ as shown in Fig.\,\ref{fig:5.7}($b$). At $r=0.9$, this homoclinic snaking occurs between $\mu= -0.44,\nu=-0.78$ and  $\mu=-0.78,\nu=-0.44$. The SN point occurs at $\mu=-0.81$ for all radii. Red dashed lines mark the snaking region as shown previously in Fig.\,\ref{fig:5.7}($a$). The localized solutions stable in the region where the trivial states stable ($\mu<0$, and $\nu<0$) and unstable in the region where the trivial states become unstable ($\mu<0$, and $\nu>0$). The green star markers refer to the LSS with an MW background and are obtained from time stepping. The $SL_s$ half line is the line where the bifurcation changes from SW to SS in the normal form  (see Fig.\,\ref{fig:5.1}). 
} 
 \label{fig:5.9}
\end{figure}

Figure \ref{fig:5.9} summarizes the results obtained from numerical continuation for radius $r=0.7$ and $r=0.9$ in the $(\nu,\mu)$-plane. The black curve refers to the extended SS and the blue curve refers to the region where LSS with the trivial state background exist at both radii. The region between the red dashed lines identifies the snaking region for radius $r=0.9$ and  the isolas for radius $r=0.7$. As we take smaller radius the snaking region becomes narrower. This also indicates that for any radius $r<0.44$ no stable LSS can occur since the bistability between two stable states lies beyond the snaking region. In Fig.\,\ref{fig:5.9} we also show points obtained from time stepping where the LSS with an MW background exist as green stars. The $SL_s$ half line is the line where the branch ends on a heteroclinic bifurcation (see Fig.\,\ref{fig:5.1}). Beyond this line, there are no SW, and time stepping evolves to a large-amplitude SS. 

\section{Localised TW} 
\label{sec:ltw}

\begin{figure}[t!]
 	\centering 
  	\includegraphics[width=0.95\linewidth]{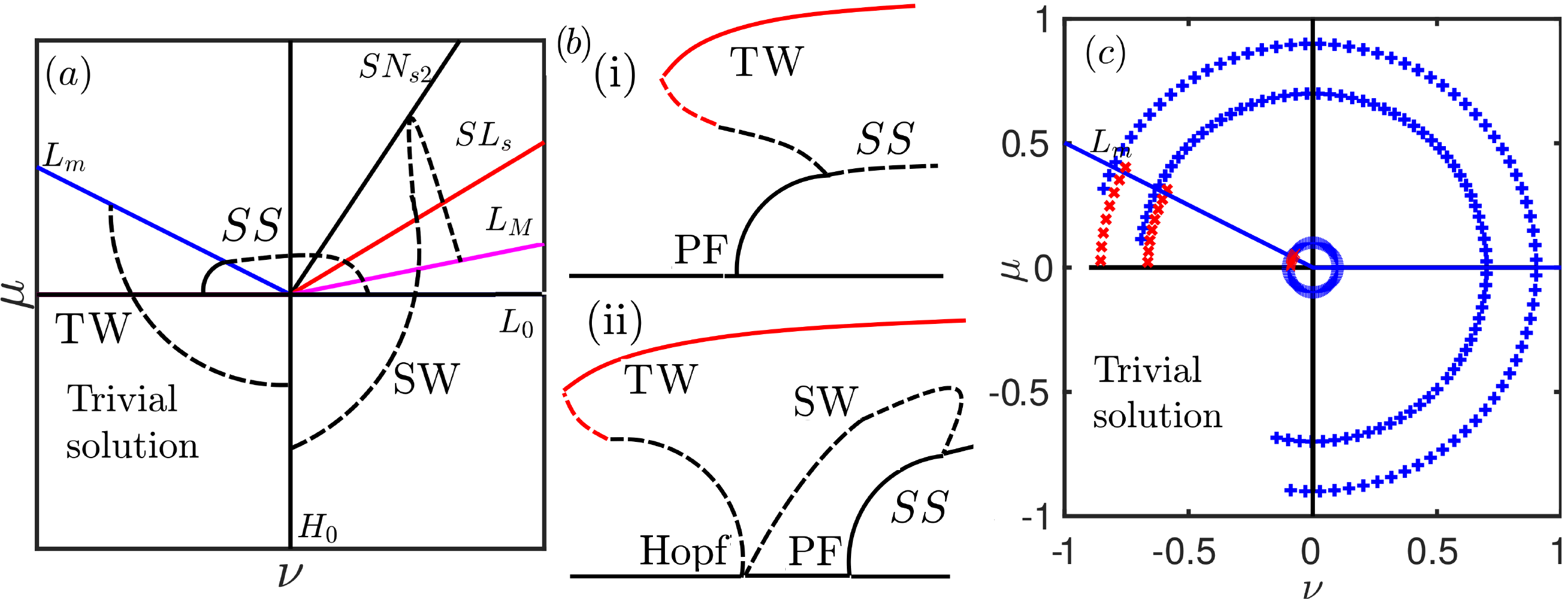}
%
 \caption{($a$) Sketch of the stability diagram for case \RN{1$-$} with $A<0$ in $(\nu,\mu)$-plane and ($b$) the corresponding bifurcation diagram from DK, where the panel ($b$)(i) represents the bifurcation above the diagonal in  the $(\nu,\mu)$-plane and ($b$)(ii) represents the bifurcation below the diagonal in the $(\nu,\mu)$-plane. ($c$) Plot of solutions obtained through time stepping Eq.\,(\ref{eq:3.19}) by time stepping with parameters values $Q_1=0.8, Q_2=0.5,C_1=-1, C_2=-0.1,C_3=-1, C_4=-0.1,C_5=-5$ and $b=2$ for radius $0.1$, $0.7$ and $0.9$. A Hopf bifurcation occurs at $\theta=270^{\circ}$ and a pitchfork bifurcation occurs at $\theta=180^{\circ}$ and $\theta=0^{\circ}$. The  blue $+$ and  red $\times$ refer to stable extended TW and SS solutions, respectively. The half line $L_m$ is the line from the normal form at which  the bifurcation from TW to SS occurs, at $\theta\approx153.3^{\circ}$. } 
    \label{fig:5.10}
\end{figure}

The second case we discuss is the one labelled as case $\RN{1$-$}$  with $A<0$ in DK where $D>0$ and $M<0$.  
Figure \ref{fig:5.10} ($a$) shows the prediction of different bifurcations and stable solutions from the TB normal form in DK. The normal form shows a stable SS occurring in the region between the pitchfork bifurcation at $\mu=0$ with $\nu<0$ and the half line $L_m$, where 
\begin{equation} 
L_m: \mu D= \nu A \qquad  A\mu<0.
\end{equation}
The TW branch bifurcates subcritically from the trivial solution at the Hopf bifurcation at $\nu=0$ with $\mu<0$. Coupling this prediction from the normal form along with the global stability requirements of the model, we expect two different localised solutions in this case: a localised travelling wave in a background of the trivial state (LTW-TS) in the third quadrant, along with the possibility of a localised travelling wave in a background of the steady state (LTW-SS) in the second quadrant. 

We consider two bifurcation scenarios to illustrate these cases, one above and one below the line $\mu=\nu$ and plot them in Fig.\,\ref{fig:5.10} ($b$). The bifurcation below the diagonal in $(\nu,\mu)$-plane (see Fig.\,\ref{fig:5.10} ($b$) (ii)) has an unstable SS branch bifurcating from a pitchfork bifurcation at $\mu=0$ with $\nu>0$. 
It also has a subcritical TW branch and unstable SW branch bifurcating from the trivial state at a Hopf bifurcation where $\nu=0, \mu<0$.  The SW branch  undergoes saddle-node (SN) bifurcation at $SN_{s2}$ and terminates at the SS branch at $L_M$. From global stability, we expect the unstable branch of TW to regain stability at a saddle-node bifurcation, giving rise to a large amplitude branch of TW solutions (as shown by red lines in Fig.\,\ref{fig:5.10} ($b$). The bifurcation above the diagonal in $(\nu,\mu)$-plane (see Fig.\,\ref{fig:5.10} ($b$)(i)) has stable SS branch bifurcating from a pitchfork bifurcation at $\mu=0$ with $\nu>0$ which becomes unstable after passing the half lines $L_m$.  The subcritical TW branch bifurcates from the SS branch at $L_m$. The trivial state is stable in the region where $\mu<0$ and $\nu<0$. This implies that we can expect coexistence between the large amplitude TW and the trivial state for values before the pitchfork bifurcation and we can expect coexistence between large amplitude TW and stable SS solutions in the range of values past the pitchfork bifurcation. 

In order to explore the fully nonlinear behaviour of the system, we run timestepping from different initial conditions for a variety of parameters at three different radii $r=0.1,0.7,0.9$ in a small (one wavelength) domain and plot the results in Fig.\,\ref{fig:5.10}($c$). In these calculations, we use $32$ grid points per wavelength. 

The trivial state is stable for the region $\nu<0$ and $\mu<0$ for all radii $r=0.1,0.7,0.9$. The small-amplitude SS bifurcate at a pitchfork bifurcation $\theta=180^{\circ}$ where $\mu=0, \nu<0$ and loses stability at $L_m$, which has the slope $\frac{\mu}{\nu}\approx -0.502$. There is a large-amplitude TW solution around the whole circle in the $(\nu,\mu)$-plane at radius $r=0.1$. At radii $r=0.7$ and $r=0.9$, the  amplitude of the TW decreases  in the region where the trivial state or the small-amplitude SS are stable and we lose the branch solutions (at potential saddle-node bifurcations). This confirms that the fully nonlinear behaviour in this case allows for bistability between two different states:  a large-amplitude TW with trivial states in the region where $\mu<0,\nu<0$ and a large-amplitude TW with small-amplitude SS in the region between the pitchfork bifurcation $\mu=0, \nu<0$ to the half line $ L_m$. Therefore, the LTW-TS and LTW-SS can be sought. 

\begin{figure}
	\centering 
  	\includegraphics[width=0.8\linewidth]{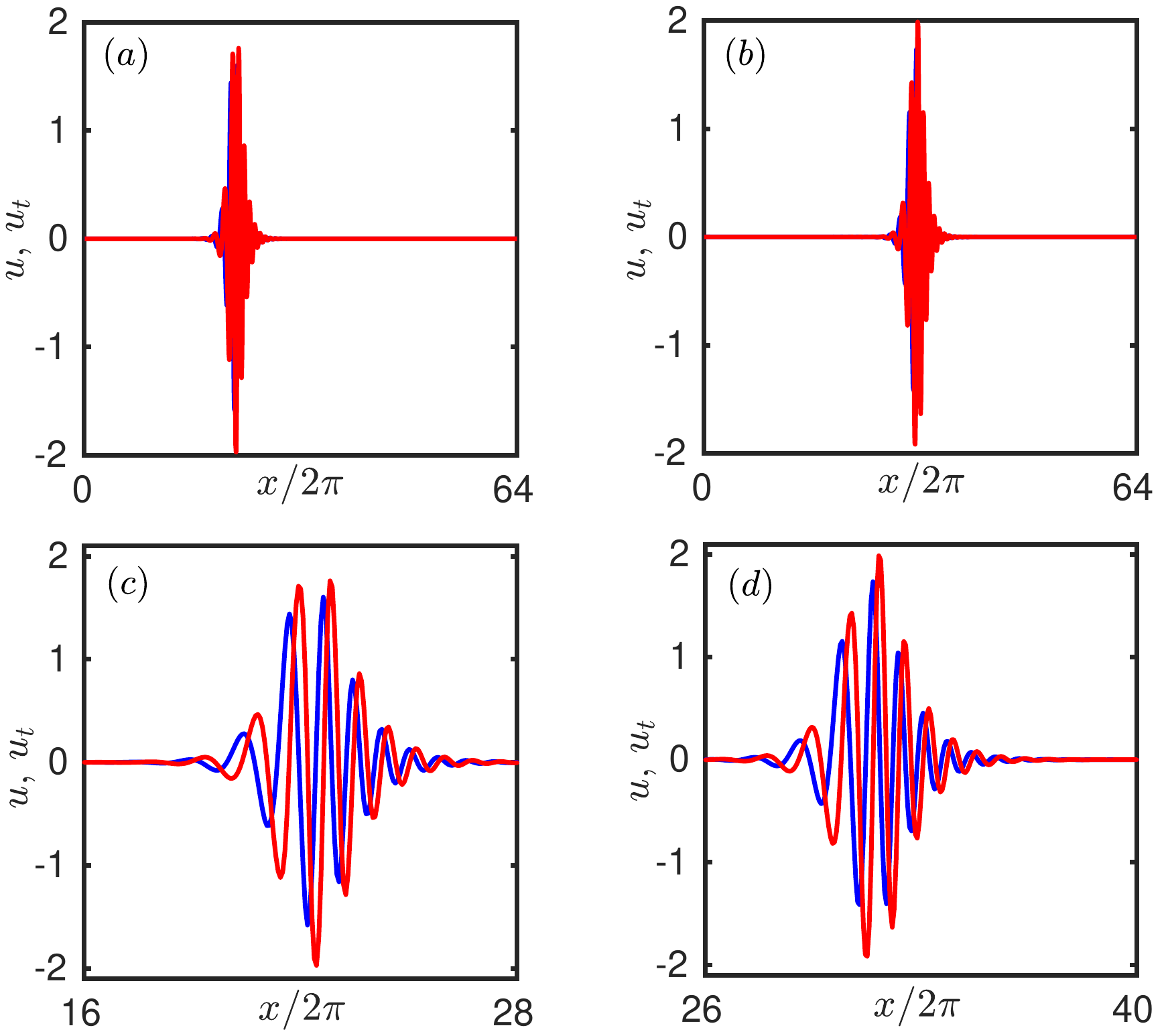}
%
         \caption{Two examples of LTW with trivial state background. ($a$) For  radius $r=0.1$ and $\theta=200^{\circ}$ and ($b$) radius $r=0.1$ and $\theta=250^{\circ}$. ($c$,$d$) Zooms of ($a$,$b$). The blue and red curves refer to $u$ and $u_t$, and the waves travel to the right. A movie of the state in ($a$) is available at \citep{SuppTBSH}.}
         \label{fig:5.12} 
         \end{figure}

\begin{figure}[t!]
	\centering 
  	\includegraphics[width=0.8\linewidth]{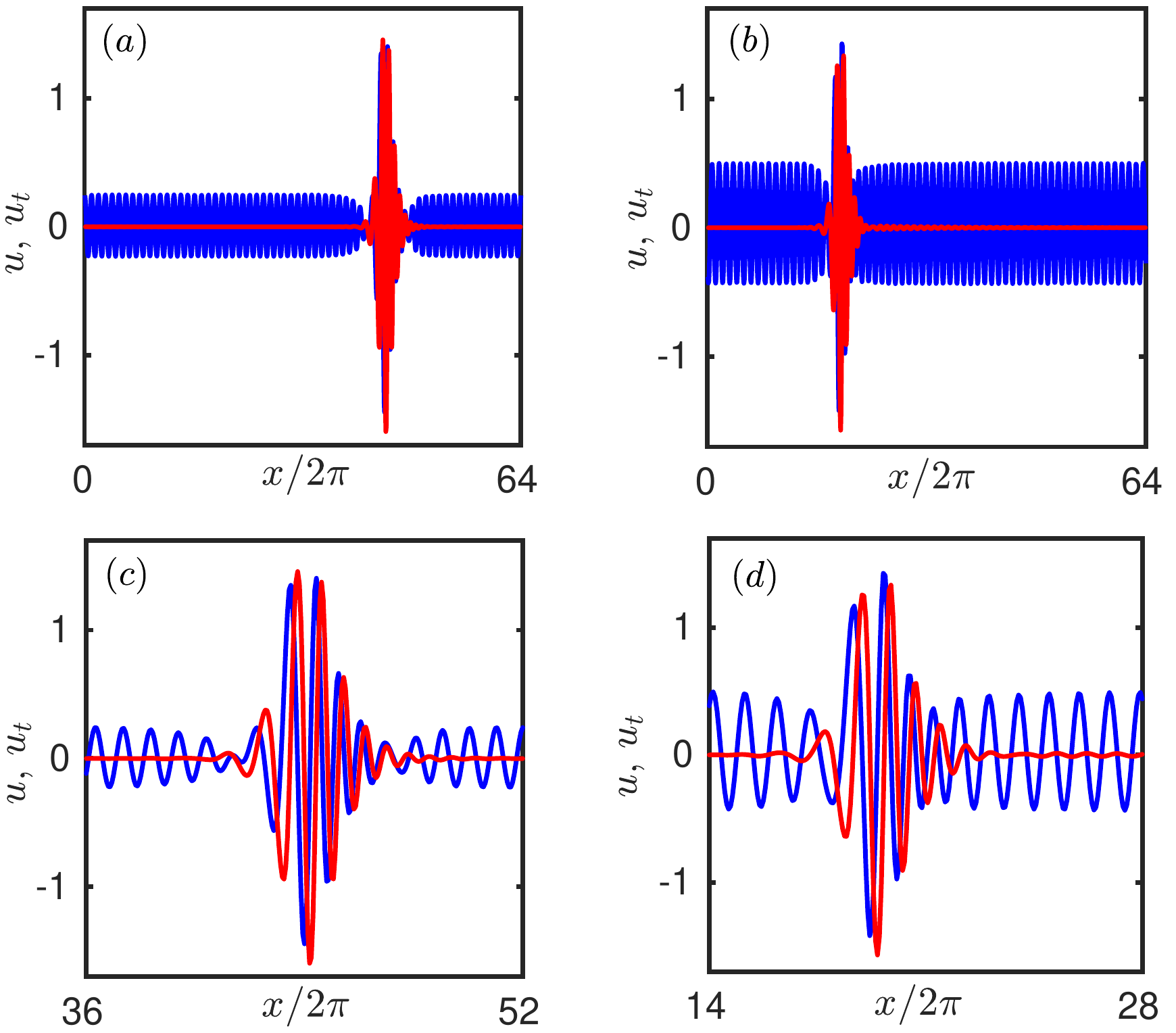}
%
 \caption{Example of LTW with SS background (a) for radius 0.1 and $\theta= 170^{\circ}$ (b) for radius 0.4 and $\theta= 160^{\circ}$ where (c,d) are zooms of (a,b). The blue and red curves refer to $u$ and $u_t$ and the wave travels to the right. A movie of the state in ($a$) is available at \citep{SuppTBSH}.}
 \label{fig:5.13} 
  \end{figure}

In order to obtain the localized state we increase the domain size to allow 64 wavelengths in the domain and perform time stepping to find an extended TW. To obtain the localized state, we use a $\sech$-envelope with different widths and do time stepping to obtain the localised state. Using this method we are able to get LTW with two different backgrounds as shown in Figs. \ref{fig:5.12} and \ref{fig:5.13}. 

  
  First, we find LTW-TS in the region where $\mu<0$ and $\nu<0$. Figure \ref{fig:5.12} shows two examples of LTW with the trivial state background for two different parameters values ($a$) for radius $r=0.1$ and $\theta= 200^{\circ}$ where $\mu=-0.034,\nu=-0.094$ and ($b$) for radius $r=0.1$ and $\theta= 250^{\circ}$ where $\mu=-0.094,\nu=-0.034$.  In these examples, the patterns of $u$ and $u_t$  move from left to right with a  group velocity smaller than the phase velocity. At  given parameter values, the LTW we find all have the same width, regardless of initial conditions (unlike LSS-TS or LSS-SW).

 Second, we find LTW-SS in the region between the pitchfork bifurcation at $\mu=0 $ where $\nu<0$ to the half line $L_m$. Figure \ref{fig:5.13} shows two examples of LTW with an SS background for ($a$) radius $r=0.1$ and $\theta= 170^{\circ}$ where $\mu=0.017, \nu=-0.098$ and ($b$) for radius $r=0.4$ and $\theta= 160^{\circ}$ where $\mu=0.14,\nu=-0.038$. The LTW move from left to right. Again, we find only LTW-SS with one chosen width in this case. 
 
 In all these LTW examples, we started with initial conditions with a wide variety of widths, but always found a localised solution with the same width, unlike in the LSS cases. 
  
\section{Conclusions}
\label{sec:conclusions}
In this paper, we have developed a simple nonlinear pattern-forming PDE model that has a Takens--Bogdanov primary bifurcation. The model is based on the Swift--Hohenberg equation, which was originally derived to describe the effects of thermal fluctuations and the evolution of roll patterns close to the onset of Rayleigh--B\'{e}nard convection and later used as a model of pattern formation in many physical problems. 
The new model can be reduced further using weakly nonlinear theory  to  the Takens--Bogdanov normal form where the pitchfork and Hopf bifurcations coincide.  The advantage of the model lies in the relative ease of investigating the nonlinear behaviour numerically and analytically, as  compared to the full PDEs of double-diffusive convection.  Alongside  the numerical results, the model is important for helping to understand the bifurcation structure and the solution behaviour close the Takens--Bogdanov point in an extended system and in particular for investigating localized solutions. 

We identified two types of localized states which have previously been observed in various systems. The first of these is the localized steady state with the trivial state as a background (LSS-TS), which was observed numerically in binary convection \citep{batiste2006spatially, mercader2009localized} and in thermosolutal convection \citep{beaume2011homoclinic}. The second localised state is that of a localized travelling wave with the trivial state as a background (LTW-TS), which was observed in binary convection \citep{niemela1990localized, barten1991localized,barten1995convection,zhao2015numerical,jung2005traveling,surko1991confined, watanabe2012spontaneous,kolodner1991collisions,kolodner1991stable,kolodner1991drift}. We have further identified two new spatially localised states: that of a localised steady state in a modulated wave background (LSS-MW) and that of a localised travelling wave in a steady state background (LTW-SS). 

To find localized states, we looked for subcritical pitchfork and Hopf bifurcations, since we expected that subsequent saddle-node bifurcations would lead to stable large-amplitude solutions coexisting with the stable trivial solutions, and possibly then to localized solutions. We identified a subcritical pitchfork bifurcation for the case \RN{4}$-$ with $A > 0$ (see Fig.\,\ref{fig:5.1}) and a subcritical Hopf bifurcation for the case \RN{1}$-$ with $A < 0$ (see Fig.  \ref{fig:5.10}). From solving the model numerically, we obtained  different types of localized states.  In case \RN{4}$-$ with $A > 0$, we obtained LSS in the region where there is  bistability between the trivial state and a branch of periodic steady states, with $\mu<0$ and $\nu<0$.  We used numerical continuation of the PDE model (\ref{eq:3.19}) to compute  the branches of localized states. The continuation method we used can  only find steady solutions, so the model is effectively the steady Swift--Hohenberg equation with solutions  depending  only on $\mu$ -- though the stabilities depend on both $\mu$ and $\nu$. 
The solutions are  associated with homoclinic connections to the trivial state, in the same manner as spatially localized solutions in the Swift--Hohenberg equation \citep{burke2006localized, burke2007snakes,burke2012localized}.  The two localized branches with odd and even numbers of peaks add an oscillation on each side as they snake back and forth until they reach the width of the domain, where they terminate on the steady state branch, at the saddle-node bifurcation (see Fig.\,\ref{fig:5.6} a and \ref{fig:5.7} i). In the absence of the saddle-node bifurcation in the periodic state where the left side of the snaking branch cannot be reached, we found isolated branches of LSS (see Fig.\,\ref{fig:5.6} ($b$) and \ref{fig:5.7} ($ii$)).  The localized solutions we obtained still exist but are unstable in the region where the trivial state becomes unstable,  where $\mu<0$ and $\nu>0$.  
 
 From time-stepping, we also found LSS with an MW background  in the region where the large-amplitude SS branch and the small-amplitude SW branch are both stable (see Fig.\,\ref{fig:5.1}, \ref{fig:5.51} ($b$)).

In case \RN{1}$-$ with $A < 0$ and from time stepping, we found LTW with the trivial state background in the region where the trivial state and a large-amplitude branch of TW are stable, with $\mu<0$ and $\nu<0$ (see Fig.\,\ref{fig:5.10}). We also found LTW with SS background in the region where the small-amplitude SS and large-amplitude TW are stable (see Fig.\,\ref{fig:5.10} and \ref{fig:5.13}). For the given parameter values, the LTW we found all have the same width, regardless of initial conditions. In contrast, LSS exist with a wide range of widths, with different numbers of peaks. In future work we will investigate why we get uniquely selected widths of LTW.   

LTW with trivial state background and with uniquely selected widths have also observed in  experimental  \citep{kolodner1994stable,kolodner1991stable,niemela1990localized,kolodner1991collisions} and numerical \citep{barten1991localized,taraut2012collisions,barten1995convection} studies of binary convection. These studies were not carried out close to the Takens--Bogdanov point, so our model does not directly apply here. Using continuation to compute the LTW solutions would need more effort due to  the time and space dependence. The numerical continuation would then require additional unknown variables: the group velocity and the temporal period. An approach to solving this problem is suggested by \cite{watanabe2011time,watanabe2012spontaneous} and we plan to undertake this in future. 

In this paper, we are interested in modelling  systems such as thermosolutal \citep{moore1991asymmetric,huppert1976nonlinear,da1981oscillations,nield1967thermohaline} and binary convection \citep{knobloch1990minimal,knobloch1986oscillatory, watanabe2012spontaneous, batiste2005simulations}, where the pitchfork and Hopf bifurcations have the same critical wavenumbers (see Fig.\,\hbox{\ref{fig:2}}), therefore, we assumed $\kcpf=\kcHopf=1$. For future investigations, if 
 $\kcpf \neq\kcHopf$ then this model could  be relevant to other problems where the pitchfork and Hopf bifurcation have different critical wavenumbers, for example in magnetoconvection \citep{chandrasekhar2013hydrodynamic,arter1983nonlinear, weiss1981convection,proctor1982magnetoconvection,MatthewsRucklidge1994,CluneKnobloch1994,DAWES2000109} and rotating convection  \citep{clune1993pattern,veronis1966motions, veronis1968large, zimmermann1988effect,DAWES2001197}. Finally, it would be interesting to explore pattern formation in this system for a wide range of parameters, both in one and two dimensions.

\section{Acknowledgements}
The authors would like to thank Cedric B\'eaume, Alan Champneys, Edgar Knobloch, David Lloyd, Jens Rademacher, Hermann Riecke and Arnd Scheel for many influential discussions. This research was supported by a PhD fellowship from Tabuk University in Saudi Arabia (H. A.), a L'Or\'eal UK and Ireland Fellowship for Women in Science (P. S.) and by the Leverhulme Trust \hbox{(RF-2018-449/9, A. M. R.)}




\end{document}